\documentclass[12pt,epsf]{article}
 \usepackage[compat=1.1.0]{tikz-feynman}
 \tikzset{graviton/.style={decorate, decoration={snake, amplitude=.4mm, segment length=1.5mm, pre length=.5mm, post length=.5mm}, double}}
 \usepackage{verbatim,graphics,graphicx,color,slashed,amsmath}
 \usepackage{bm}
 \usepackage{cite}
 \usepackage{amssymb}
 \usepackage{braket}
 \usepackage{ascmac}
 \usepackage{multirow}
\usepackage[]{hyperref}
 \hypersetup{colorlinks,bookmarks,unicode,linktocpage=true,linkcolor=blue, anchorcolor=blue, citecolor=blue}
 \usepackage{comment} 
 \usepackage{ulem}
 
\usepackage[hang,small,bf]{caption}
\usepackage[subrefformat=parens]{subcaption}
\newcounter{num}

\begin{document}
\thispagestyle{empty}

\vspace*{15mm}
\begin{center}
{\Large\bf Gravitational Positivity Bounds \\on Higgs-Portal Dark Matter}
\vspace{7mm}

\baselineskip 18pt
{\bf Kimiko Yamashita}
\vspace{2mm}

{\it
Department of Physics, Ibaraki University, Mito 310-8512, Japan
\newline
kimiko.yamashita.nd93@vc.ibaraki.ac.jp}\\
\vspace{10mm}
\end{center}
\begin{center}
\begin{minipage}{14cm}
\baselineskip 16pt
\noindent
\begin{abstract}
Gravitational positivity bounds are constraints on a renormalizable theory in the presence of a massless graviton, under the assumption that the gravitational theory is ultraviolet-completed by a perturbative string theory.
We derive these bounds for the Higgs-portal scalar dark matter model using the forward scattering process $\phi \phi \to \phi \phi$. 
We find that, in the absence of a dark matter self-coupling, new physics beyond the Higgs-portal dark matter interaction
must appear below an energy scale of $10^{10}\,\mathrm{GeV}$
if the dark matter mass is smaller than the Higgs boson mass.
We further find that, in the presence of both interactions,
achieving a cutoff scale at the grand unified theory scale
generally requires a dark matter mass of order
$10^{10}$--$10^{11}\,\mathrm{GeV}$ (or above),
with larger values favored when the four-point self-coupling
plays a significant role.
For such heavy Higgs-portal dark matter,
the observed relic abundance of dark matter in the Universe
can be successfully reproduced via the freeze-in mechanism
with a tiny Higgs-portal coupling,
$\lambda_{h\phi} \lesssim 3.5 \times 10^{-11}$.
The reheating temperature is then constrained to be
$T_{\mathrm{reh}} \lesssim 10^{14}\,\mathrm{GeV}$
by the positivity bounds on the dark matter mass, for $m_\phi$ up to $10^{15}$~GeV, 
an order of magnitude below the grand unified theory scale.
\end{abstract}
\end{minipage}
\end{center}

\baselineskip 18pt
\def\thefootnote{\fnsymbol{footnote}}
\setcounter{footnote}{0}
\newpage

\section{Introduction}\label{sec:introduction}
The existence of dark matter (DM) is one of the primary motivations for considering physics beyond the Standard Model (SM).
If we assume the existence of a SM gauge-singlet scalar particle, it can interact with the Higgs boson through a renormalizable Lagrangian.
When this scalar field is odd under a $Z_2$ symmetry, whereas all SM particles are even, the scalar is stable and does not decay into SM particles.
As a result, the scalar boson becomes a viable DM candidate, and its interaction with all other particles proceeds through the so-called Higgs portal.
This Higgs-portal scalar DM model is a minimal and well-studied extension of the SM with rich phenomenology~\cite{Silveira:1985rk,McDonald:1993ex,Burgess:2000yq} (see also~\cite{Arcadi:2019lka,Lebedev:2021xey} for reviews and \cite{Kim:2023pwf,Kim:2023bbs,Yamashita:2024krp,DeFelice:2025qva} for studies including higher-dimensional operators).

The viability of such DM models can be evaluated through their ability to reproduce the observed DM relic abundance 
in the Universe, and through their consistency with experimental and observational constraints.
In addition to these phenomenological approaches, the models can be examined from the perspective of ultraviolet (UV) completion, taking into account the fundamental principles of unitarity, analyticity, and Lorentz invariance of the $S$-matrix.

A powerful tool in this regard is the use of positivity bounds~\cite{Adams:2006sv,Pham:1985cr,Ananthanarayan:1994hf}.
The original positivity bounds were derived for higher-dimensional operators of dimension-8 and above.
Dimension-8 operators are accompanied by the fourth power of the cutoff scale in the denominator,
so they are suppressed compared to dimension-6 and dimension-5 operators, or the unsuppressed renormalizable interactions. 
However, by introducing observables that are insensitive to dimension-6 operators, one can isolate the effects of dimension-8 interactions and explore the imposed positivity bounds and their possible violations~\cite{Li:2022rag}.
Positivity bounds on Higgs-portal scalar DM with local dimension-8 operators have been investigated by Kim, Lee, and the author in~\cite{Kim:2023pwf,Kim:2023bbs}.

Constraints can also be placed on renormalizable interactions by exploiting gravitational positivity bounds~\cite{Tokuda:2020mlf}.
Assuming that gravity is UV-completed by a weakly coupled string theory, the renormalizable theory must satisfy bounds imposed by this UV gravitational completion.
These constraints are referred to as gravitational positivity bounds~\cite{Tokuda:2020mlf} (see also~\cite{Hamada:2018dde} for related arguments involving Regge states coupled to the graviton, and~\cite{Aoki:2021ckh,Noumi:2021uuv,Noumi:2022zht,Aoki:2023khq,Kim:2024iud} for applications to the SM and phenomenological models).

In this paper, we derive gravitational positivity bounds for the renormalizable Higgs-portal scalar DM interactions and discuss their phenomenological implications.
The remainder of this paper is organized as follows.
In Section~\ref{sec:model}, we introduce the Lagrangian of the model.
In Section~\ref{sec:posi}, we review the derivation of gravitational positivity bounds for renormalizable interactions and present the bounds obtained for the Higgs-portal scalar DM model.
In Section~\ref{sec:dm_pheno}, we discuss the phenomenological implications
of these bounds.
Finally, in Section~\ref{sec:summary},
we summarize our findings and outline directions for future work.

\section{Higgs-Portal Scalar Dark Matter}\label{sec:model}

We consider DM to be a real scalar field $\phi$, which is a SM gauge singlet and odd under a discrete $Z_{2}$ (dark) parity, i.e., $\phi \to -\phi$, whereas all SM particles are even under this parity.
We assume that the Lagrangian is symmetric under the $Z_{2}$ transformation to ensure DM stability, so that a DM particle $\phi$ cannot decay into SM particles.  
The renormalizable interaction of DM with the SM sector is of the Higgs-portal type, meaning that the interactions occur through the Higgs boson.  
The Lagrangian for the Higgs-portal scalar DM field $\phi$ is given by~\cite{Kanemura:2010sh,Beniwal:2015sdl}
\begin{align}
\mathcal{L} = \mathcal{L}_{\mathrm{SM}} + \frac{1}{2}\partial_{\mu}\phi\partial^{\mu}\phi -\frac{1}{2}\mu^2_{\phi}\phi^2-\frac{1}{4!}\lambda_{\phi}\phi^4 -\frac{1}{2}\lambda_{h\phi}H^{\dagger}H\phi^2, \label{eq:lag}
\end{align}
where $ \mathcal{L}_{\mathrm{SM}}$ is the SM Lagrangian and $H$ is the SM Higgs doublet.
The bare mass term of the DM field is allowed, whereas linear and cubic terms such as $\phi$, $\phi^3$, and $H^{\dagger} H \phi$ are forbidden by the $Z_2$ symmetry.

The SM Higgs doublet acquires a vacuum expectation value (VEV) $v$, leading to spontaneous electroweak symmetry breaking. In the unitary gauge, the Higgs doublet is written as
\begin{align}
H =\frac{1}{\sqrt{2}}\begin{pmatrix}
0 \\
v + h
\end{pmatrix},
\label{eq:sb}
\end{align}
where $h$ denotes the physical SM Higgs field.
Substituting Eq.~\eqref{eq:sb} into Eq.~\eqref{eq:lag}, we obtain
\begin{align}
\mathcal{L} = \mathcal{L}_{\mathrm{SM}} + \frac{1}{2}\partial_{\mu}\phi\partial^{\mu}\phi -\frac{1}{2}m^2_{\phi}\phi^2
-\frac{1}{4!}\lambda_{\phi}\phi^4 -\frac{1}{2}\lambda_{h\phi}v h\phi^2-\frac{1}{4}\lambda_{h\phi}h^2\phi^2, \label{eq:lag2}
\end{align}
with the physical DM mass given by
\begin{align}
m^2_{\phi} = \mu^2_{\phi} +\frac{1}{2}\lambda_{h\phi} v^2. \label{eq:dm_mass}
\end{align}
This model is characterized by three independent parameters: the scalar DM mass (expressed in terms of the Lagrangian parameters via Eq.~\eqref{eq:dm_mass}), the Higgs-portal coupling, and the DM self-coupling, which we denote by
\begin{align}
\{m_{\phi}, \lambda_{h\phi}, \lambda_{\phi}\}.
\end{align}

\section{Gravitational Positivity Bounds and Application to Higgs-Portal Scalar DM} \label{sec:posi}

Gravitational positivity bounds~\cite{Tokuda:2020mlf} can be applied to renormalizable interactions.  
In this section, we first review the framework of gravitational positivity bounds.  
Following this review, in Sec.~\ref{sec:posi_bound}, we derive the gravitational positivity bound for the $\phi \phi \to \phi \phi$ process using the Higgs-portal scalar DM Lagrangian given in Eq.~\eqref{eq:lag2}.

\subsection{Pedagogical Review}\label{sec:posi_review}

We review the derivation of gravitational positivity bounds 
using the tree-level string-state contributions to the $\phi \phi \to \phi \phi$ scattering process.
%
%
Although we explicitly consider the $\phi \phi \to \phi \phi$ process as an example, the discussion is generic.
We introduce two scales $\Lambda < \Lambda_{\text{QG}}$ above the electroweak scale,
the new physics scale $\Lambda$ and the quantum gravity scale $\Lambda_{\text{QG}}$, and assume the following:
\begin{itemize}
  \item[\textbf{(i)}] 
At very high energies above the scale $\Lambda_{\text{QG}}$, gravity is UV completed by string theory.  
$\Lambda_{\mathrm{QG}}$ is identified with the string scale.  
An infinite tower of massive higher-spin states, i.e., a Regge tower, as predicted in perturbative string theory, becomes relevant for scattering amplitudes. 
We consider only the tree-level contributions of Regge (string) states to the scattering process.

The scattering amplitude $\mathcal{M}(s, t)$ in the complex Mandelstam $s$-plane 
is assumed to satisfy
\begin{align}
\lim_{|s| \to \infty} \left| \frac{\mathcal{M}(s, t < 0)}{s^2} \right| = 0,
\label{eq:amp_s_limit}
\end{align}
where $t = 0$ is excluded to avoid the $t$-channel graviton pole.  
This behavior in Eq.~\eqref{eq:amp_s_limit} is not a strong assumption, as it is satisfied by the Froissart-Martin bound~\cite{Froissart:1961ux,Martin:1962rt}, which follows from the analyticity (i.e., causality) of the amplitude.  
Moreover, it also arises in tree-level string theory~\cite{Tokuda:2020mlf,Hamada:2023cyt}.

\item[\textbf{(ii)}] 
At energies below the string scale $\Lambda_{\text{QG}}$ but above the new physics scale $\Lambda$, 
physics is described by a graviton coupled to a quantum field theory (QFT) describing particles and their interactions. 
The QFT is assumed to respect Lorentz symmetry, unitarity, and analyticity as usual.
Apart from this requirement, a concrete Lagrangian is not needed.
In particular, in addition to the SM and the DM particle $\phi$, various new physics contributions, 
i.e. particles with masses larger than $\Lambda$ and their interactions, may exist but need not be specified.

\item[\textbf{(iii)}] 
At energies below the new physics scale $\Lambda$, only the SM and the DM particle $\phi$ are present
in the effective field theory (EFT).
The renormalizable part of the effective Lagrangian is Eq.~\eqref{eq:lag2}.
EFT operators of dimension $n > 4$, suppressed by $\Lambda^{n-4}$, arise from integrating out the heavy states (and their interactions) that are assumed to exist above $\Lambda$ (see item (ii)).  
Ordinary gravitational interactions, i.e., interactions with a massless graviton, are included.  
These are described by dimension-5, Planck-suppressed, tree-level interactions between the massless graviton and all particles via their energy-momentum tensor.  
Loop-level interactions of the massless graviton, which are further suppressed by the Planck scale, are neglected
\end{itemize}

We consider the 2-to-2 elastic scattering amplitude 
$\mathcal{M}(\phi(p_1) \phi(p_2) \to \phi(p_3) \phi(p_4))$   
and assume that the Mandelstam variable $t = (p_3 - p_1)^2$ satisfies the condition $0 < -t \ll 4 m_\phi^2$.
The upper bound facilitates the derivation of the dispersion relations we will be using in our analysis.
See Appendix A.
The condition $0<-t$ is required to avoid the $t$-channel pole arising from the massless graviton. 
Eventually, we take the limit $t \to -0$.

By considering the extended (almost) forward elastic scattering amplitude $\mathcal{M}(s, t)$ in the complex $s$-plane, one can relate the low-energy elastic forward amplitudes to the imaginary part of the amplitude at high energy.  
For details of the derivation, see Appendix~\ref{sec:appendix_posi}, especially the steps from Eq.~\eqref{eq:amp1_s} to Eq.~\eqref{eq:amp2_s}.

The amplitude in the $s$-plane can be written as (Appendix~\ref{sec:appendix_posi})
\begin{align}
\mathcal{M}(s, t) = \left(s - 2 m_\phi^2 + \frac{t}{2}\right)^2 
\oint_{\mathcal{C}} \frac{ds'}{2\pi i} \frac{\mathcal{M}(s', t)}{(s' - s) \left(s' - 2 m_\phi^2 + t/2\right)^2}.
\label{eq:amp1_s}
\end{align}
The twice-subtracted dispersion relation is (Appendix~\ref{sec:appendix_posi})
\begin{align}
\mathcal{M}(s, t) &= \left[\frac{a_{-1}(t)}{s - m_\phi^2} + (s \leftrightarrow u(s,t))\right] + a_0 + a_1(t) s \nonumber \\ 
&\quad + \frac{2 \left(\bar{s} + \bar{t}/2\right)^2}{\pi} 
\int_{4 m_\phi^2}^{\infty} d\mu \frac{\mathrm{Im} \mathcal{M}(\mu + i\epsilon, t)}
{\left(\bar{\mu} + \bar{t}/2\right) \left[\left(\bar{\mu} + \bar{t}/2\right)^2 - \left(\bar{s} + \bar{t}/2\right)^2\right]},
\label{eq:amp2_s}
\end{align}
where the barred variables are defined as $\bar{z} = z - 4 m_\phi^2 / 3$.  
The explicit forms of the functions $a_{-1}(t)$, $a_0$, and $a_1(t)$ are not relevant for our discussion here.

We note that the pole term $a_{-1}(t)/(s-m_\phi^2) + (s\leftrightarrow u(s,t))$ in
Eq.~\eqref{eq:amp2_s} is generically expected in the dispersion relation for a scalar
field with a cubic self-interaction, corresponding to on-shell single-particle exchange in the
$s$- and $u$-channels. Under the $Z_2$ symmetry imposed in Sec.~\ref{sec:model},
however, such odd-point interactions (e.g., $\phi^3$) are forbidden, so
this pole is in fact absent for the Higgs-portal dark matter model
considered here. We retain it in this section as part of a general,
pedagogical derivation valid for an arbitrary scalar potential. Since
the pole term does not contribute to the coefficient of $s^2$ in the
low-energy amplitude (cf.\ Eq.~\eqref{eq:amp3}), which is the only piece entering
the positivity bound derived in Sec.~\ref{sec:posi_bound}, its presence or absence in
this review does not affect any of our results.
\begin{figure}[!t]
\begin{minipage}[b]{0.24\linewidth}
\includegraphics[width=\textwidth,clip]{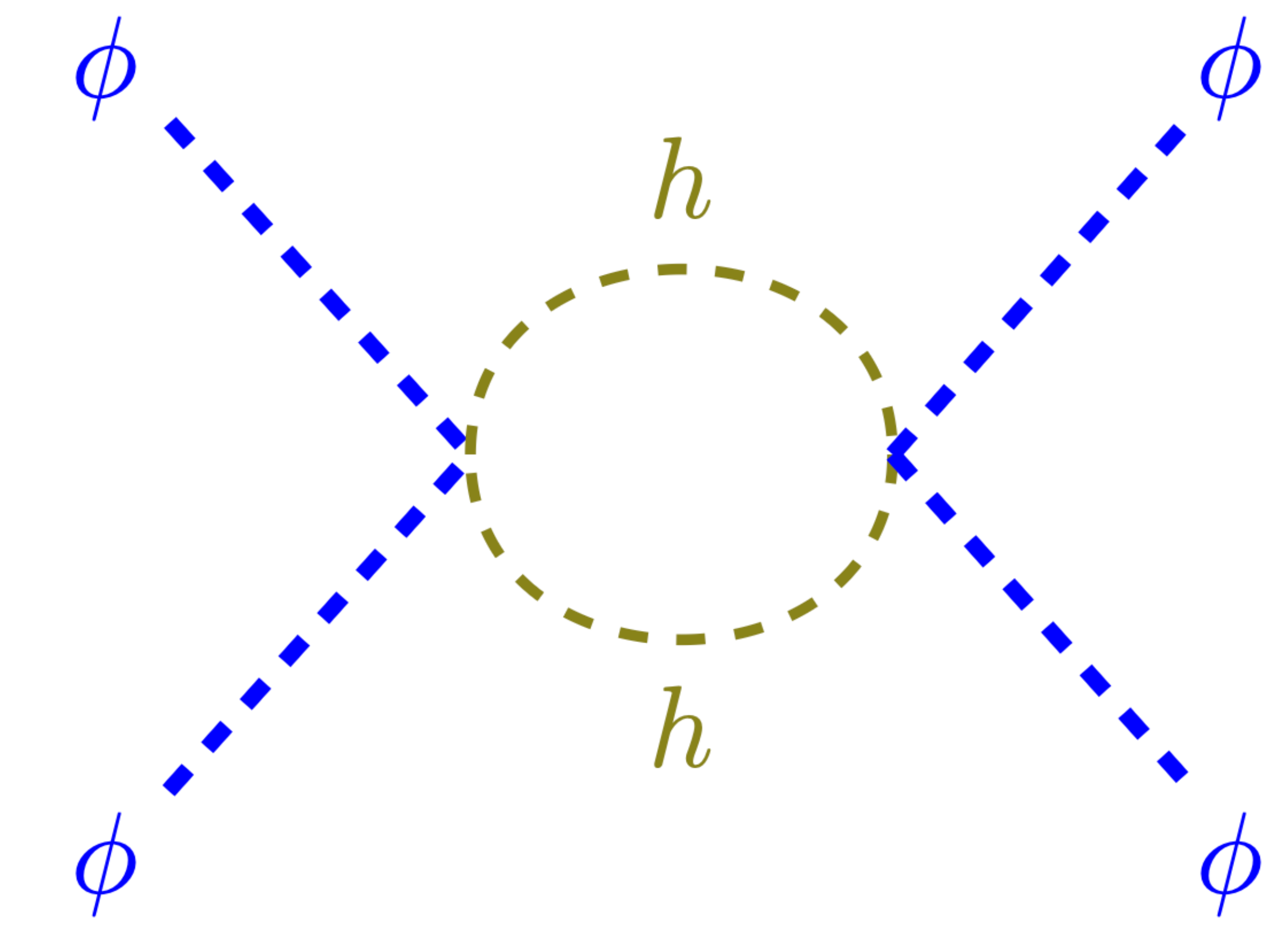}
\subcaption{first: $\tilde{a}_2(t) s^2$}
\end{minipage}\hspace{2pt}
\begin{minipage}[b]{0.24\linewidth}
\centering
\includegraphics[width=\textwidth,clip]{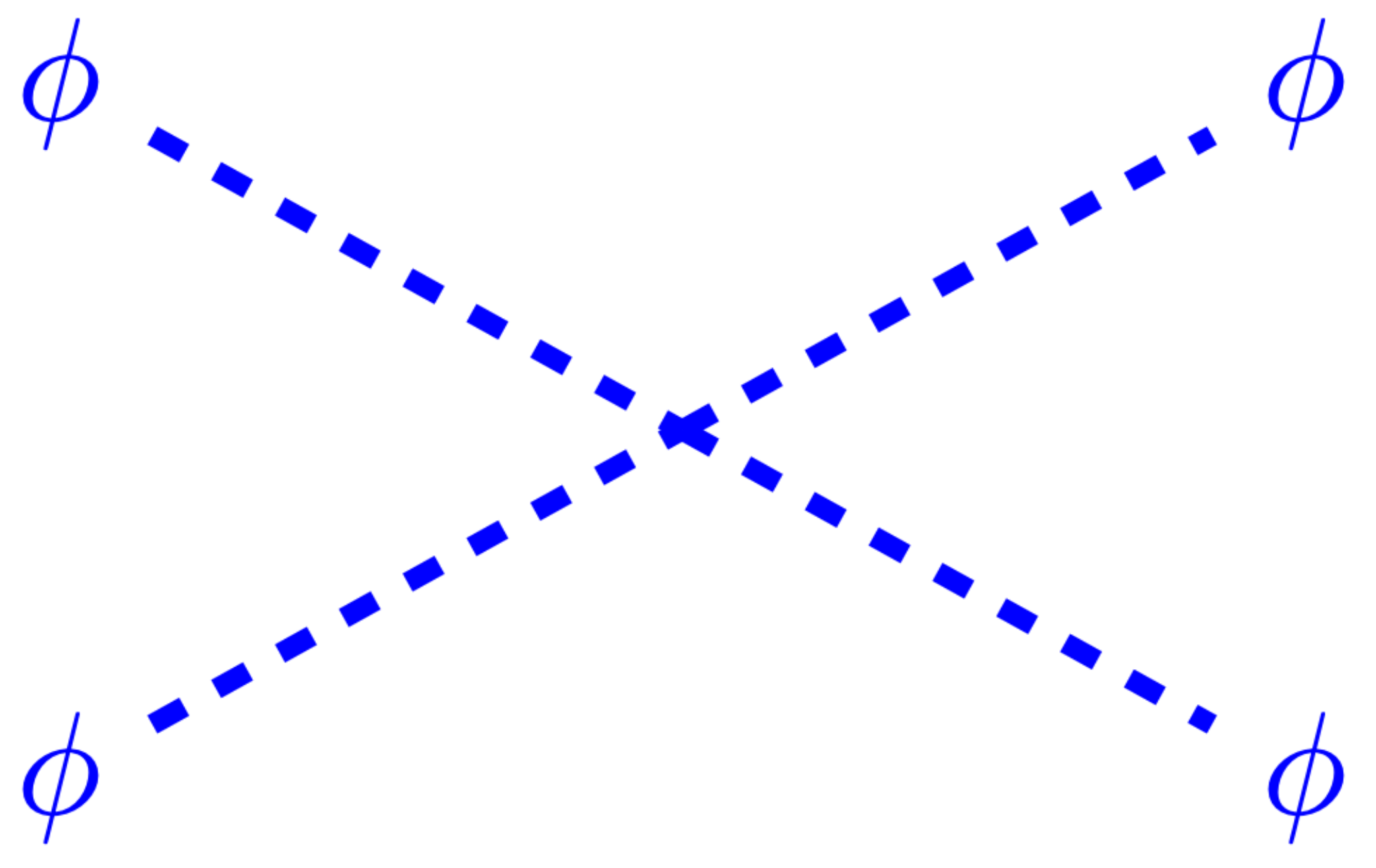}
\subcaption{first: $\tilde{a}_2(t) s^2$}
\end{minipage}\hspace{2pt}
\begin{minipage}[b]{0.24\linewidth}
\centering
\begin{tikzpicture}
\begin{feynman}
\vertex (a1) {\(\textcolor{blue}{\phi} \)};
\vertex at (1.2,0)(a2);
\vertex at (2.4,0)(a3);
\vertex at (3.6,0)(a4) {\(\textcolor{blue}{\phi} \)};
\vertex at (1.8,-1.0)(d);
\vertex at (0,-2.0) (c1){\(\textcolor{blue}{\phi} \)};
\vertex at (3.6,-2.0) (c3) {\(\textcolor{blue}{\phi} \)};
\vertex at (1.8,-2.0) (c2);
\diagram* {
(a1) -- [scalar, line width=0.6mm, color=blue] (a2) -- [scalar,  line width=0.6mm, color=blue, edge label=\(\textcolor{blue}{\phi} \)] (a3) -- [scalar, line width=0.6mm, color=blue] (a4),
(a2) -- [scalar, edge label'=\(h\),  line width=0.4mm, color=olive] (d) -- [scalar, edge label'=\(h\),  line width=0.4mm, color=olive] (a3),
(c1) -- [scalar, line width=0.6mm, color=blue] (c2)-- [scalar, line width=0.6mm, color=blue](c3),
(d) -- [graviton, edge label=\(G\), color=purple] (c2),
};
\end{feynman}
\end{tikzpicture}
\subcaption{second: $-g_2(t)s^2/M^2_{\text{pl}}$}
\end{minipage}
\begin{minipage}[b]{0.24\linewidth}
\centering
\begin{tikzpicture}
\begin{feynman}
\vertex (a1) {\(\textcolor{blue}{\phi} \)};
\vertex at (1.8,0)(a2);
\vertex at (3.6,0)(a4) {\(\textcolor{blue}{\phi} \)};
\vertex at (0,-2.0) (c1){\(\textcolor{blue}{\phi} \)};
\vertex at (3.6,-2.0) (c3) {\(\textcolor{blue}{\phi} \)};
\vertex at (1.8,-2.0) (c2);
\diagram* {
(a1) -- [scalar, line width=0.6mm, color=blue] (a2) -- [scalar, line width=0.6mm, color=blue](a4),
(c1) -- [scalar, line width=0.6mm, color=blue] (c2) -- [scalar, line width=0.6mm, color=blue](c3),
(a2) -- [graviton, edge label=\(G\), color=purple] (c2),
};
\end{feynman}
\end{tikzpicture}
\subcaption{last: $-Cs^2/(M^2_\text{pl} t)$}
\end{minipage}\hspace{5pt}
\caption{Representative contributions to the low-energy $\phi \phi \to \phi \phi$ scattering amplitude.
“First,” “second,” and “last” in the subcaptions refer to the corresponding terms in Eq.~\eqref{eq:amp3}.}
\label{fig:low_energy}
\end{figure}

From assumption (iii), the terms proportional to $s^2$ in the low-energy amplitude ($E < \Lambda$) are
\begin{align}
\mathcal{M}(s, t) \supset a_2(t) s^2
= \tilde{a}_2(t) s^2 - g_2(t)s^2/M^2_{\text{pl}} - C\frac{s^2}{M^2_\text{pl} t}, \label{eq:amp3}
\end{align}
where we neglect terms not proportional to $s^2$.
(Including them will restore crossing symmetry.)
As shown in Figure~\ref{fig:low_energy}, the contributions on the right-hand side of Eq.~\eqref{eq:amp3} originate from four types of 
diagrams (a)--(d):
\begin{itemize}
\item[(a)] The first term receives loop-level contributions from the Higgs-portal DM.
\item[(b)] The first term also receives tree-level contributions from dimension-8 derivative operators, such as $(\partial \phi)^4/\Lambda^4$.
\item[(c)] The second term corresponds to a $t$-channel graviton contribution with a loop of SM and $\phi$ particles.
\item[(d)] The last term arises from a tree-level $t$-channel graviton exchange, with a positive constant $C$ and the Planck mass $M_{\text{pl}}$.
\end{itemize}
For cases (a) and (c), the full contributions of diagrams involving the Higgs-portal DM are discussed in Sec.~\ref{sec:posi_bound}.
The last term, i.e., the $t$-channel contribution from a massless graviton, $\mathcal{M}(s, t) \sim - Cs^2/(M^2_{\text{pl}}t)$, diverges in the forward limit ($t \to -0$). Using $u = 4 m_\phi^2 - s - t$ and 
taking of the limit $t \to -0$ at the end of the derivation, only the Mandelstam $s$ contribution remains.

A term proportional to $s^2$ arises only from the final term of the dispersion relation in Eq.~\eqref{eq:amp2_s}:
\begin{align}
\mathcal{M}(s, t) \supset 2\frac{\left(\bar{s} + \bar{t}/2\right)^2}{\pi} \int_{4m_\phi^2}^{\infty} d\mu \frac{\mathrm{Im}\mathcal{M}(\mu + i\epsilon, t)}{(\bar{\mu} + \bar{t}/2) \left[(\bar{\mu} + \bar{t}/2)^2 - (\bar{s} + \bar{t}/2)^2 \right]}. \label{eq:amp2_s2}
\end{align}
From Eqs.~\eqref{eq:amp3} and \eqref{eq:amp2_s2}, by taking the second derivative of the amplitude with respect to $s$ at $s = 2 m_\phi^2 - t/2$, we obtain
\begin{align}
2\tilde{a}_2(t) - 2g_2(t)/M^2_{\text{pl}} - 2\frac{C}{M^2_\text{pl} t} &= \frac{4}{\pi} \int_{4 m_\phi^2}^{\infty} d\mu \frac{\mathrm{Im} \mathcal{M}(\mu + i \epsilon, t)}{(\mu - 2 m_\phi^2 + t/2)^3}.
\label{eq:amp_diffs0}
\end{align}
Here we include the $u$-channel contributions in the LHS of Eq.~\eqref{eq:amp_diffs0} via 
$u = 4 m_\phi^{2} - s - t \sim 4 m_\phi^{2} - s$, 
which are absorbed into $\tilde{a}_2(t)$ and $g_2(t)$, respectively.
The right-hand side (RHS) of Eq.~\eqref{eq:amp_diffs0} can be divided into three energy regions:
\begin{align}
(\text{RHS of Eq.~\eqref{eq:amp_diffs0}}) = \frac{4}{\pi} \left( \int_{4 m_\phi^2}^{\Lambda^2} + \int_{\Lambda^2}^{\Lambda_\mathrm{QG}^2} + \int_{\Lambda_\mathrm{QG}^2}^{\infty} \right) d\mu \frac{\mathrm{Im} \mathcal{M}(\mu + i \epsilon, t)}{(\mu - 2 m_\phi^2 + t/2)^3}.
\label{eq:amp_diffs1}
\end{align}
Following the analysis of Ref.~\cite{Bellazzini:2016xrt,deRham:2017imi,deRham:2017xox}
we rearrange terms of Eq.~\eqref{eq:amp_diffs0} into the form
\begin{eqnarray}
\lefteqn{
2 \tilde{a}_2(t) - \frac{4}{\pi} \int_{4 m_\phi^2}^{\Lambda^2} d\mu \frac{\mathrm{Im} \mathcal{M}(\mu + i \epsilon, t)}{(\mu - 2 m_\phi^2 + t/2)^3} - 2 g_2(t)/M^2_{\text{pl}} 
} \vphantom{\Bigg|}\cr
& = & 2 \frac{C}{M^2_\text{pl} t} + \frac{4}{\pi} \int_{\Lambda_\mathrm{QG}^2}^{\infty} d\mu \frac{\mathrm{Im} \mathcal{M}(\mu + i \epsilon, t)}{(\mu - 2 m_\phi^2 + t/2)^3} + \frac{4}{\pi} \int_{\Lambda^2}^{\Lambda_\mathrm{QG}^2} d\mu \frac{\mathrm{Im} \mathcal{M}(\mu + i \epsilon, t)}{(\mu - 2 m_\phi^2 + t/2)^3}.
\vphantom{\Bigg|}
\label{eq:amp_new}
\end{eqnarray}
In the forward limit $t \to -0$, the first term on the right-hand side of Eq.~\eqref{eq:amp_new} diverges to $-\infty$ since $C>0$.
However, this divergence is canceled by the contributions of tree-level Regge states in the first integral, corresponding to energies $E > \Lambda_\mathrm{QG}$.
Above the string or quantum gravity scale, $\Lambda_{\mathrm{QG}}=\alpha'^{-1/2}$, 
a tower of massive higher spin states (a Regge tower) appear, as assumed in assumption~(i). 
The scattering amplitude receives contributions from these Regge states~\cite{Tokuda:2020mlf}.
Here the scale $\alpha'$ corresponds to the usual $\alpha'$ parameter in string theory.
The associated divergence is almost canceled by summing the contributions of the Regge states, leaving at most a possible remnant of order $\mathcal{O}(\alpha'/M^2_{\mathrm{pl}}) = \mathcal{O}(1/(\Lambda_{\mathrm{QG}}^2 M^2_{\mathrm{pl}}))$,
as shown by Tokuda et al.~\cite{Tokuda:2020mlf} using string theory. (See also Refs.~\cite{Aoki:2023khq,Hamada:2023cyt}.)

Now, Eq.~\eqref{eq:amp_new} can be rewritten by neglecting all string states except for the Regge states and by omitting loop-level contributions in the UV region, $E > \Lambda_{\mathrm{QG}}$:
\begin{align}
& 2\tilde{a}_2(t \to -0)
 - \frac{4}{\pi} \int_{4m_\phi^{2}}^{\Lambda^{2}} d\mu
   \frac{\mathrm{Im}\,\mathcal{M}(\mu + i\epsilon, t \to -0)}
        {(\mu - 2 m_\phi^{2})^{3}}
 - \frac{2 g_2(t \to -0)}{M_{\mathrm{pl}}^{2}}
 \nonumber \\
& \qquad =
   \frac{4}{\pi} \int_{\Lambda^{2}}^{\Lambda_{\mathrm{QG}}^{2}} d\mu
   \frac{\mathrm{Im}\,\mathcal{M}(\mu + i\epsilon, t \to -0)}
        {(\mu - 2 m_\phi^{2})^{3}}
   > 0 .
\label{eq:amp_new2}
\end{align}
Here we take the forward limit $t \to -0$.  
Note that the right-hand side of Eq.~\eqref{eq:amp_new2} is manifestly positive due to unitarity, assumption (ii),
since unitarity leads to the optical theorem for forward 2-to-2 elastic scattering: 
\begin{align}
\mathrm{Im}\, \mathcal{M}(p_1, p_2 \to p_1, p_2) = s\, \sigma_\mathrm{tot}(p_1, p_2 \to \text{anything}) > 0 \quad (\text{in the massless limit}).
\end{align}
Note that the denominator of the integrand on the right-hand side of Eq.~\eqref{eq:amp_new2} is also positive as $\Lambda \gg m_\phi$ at the lower limit of the integral.
This implies that the left-hand side of Eq.~\eqref{eq:amp_new2} is positive, despite the second and third terms contributing with minus signs.

Therefore, Eq.~\eqref{eq:amp_new2} can be written as
\begin{align}
B(\Lambda) &= B_{\text{non-grav}} + B_{\text{grav}} > 0, \label{eq:grav_positive} \\
B_{\text{non-grav}} &= 2 \tilde{a}_2(t \to -0) - \frac{4}{\pi} \int_{4m^2_\phi}^{\Lambda^2} d\mu \frac{\mathrm{Im}\, \mathcal{M}(\mu + i\epsilon, t\to-0)}{(\mu - 2 m^2_\phi)^3}, \label{eq:non-grav} \\
B_{\text{grav}} &= - \frac{2 g_2(t \to -0)}{M^2_{\text{pl}}} = \lim_{t \to -0} \left(\frac{\partial^2 \mathcal{M}_{\text{grav}}(s,t)}{\partial s^2}\right)_{s=2m^2_\phi}.
\end{align}
Here $B_{\text{grav}}$ represents the subtraction of the $s^{2}$ dependence of the amplitude, so its mass dimension is $-4$.
Non-gravitational part, $B_{\text{non-grav}}$, likewise has mass dimension $-4$, and it involves the subtraction of a positive contribution from the integrand.

Neglecting the graviton contributions (i.e., terms suppressed by $1/M^2_{\text{pl}}$, corresponding to $B_{\text{grav}}$) and using Eq.~\eqref{eq:amp_new}, $B_{\text{non-grav}}$ can be expressed by ignoring the UV details of the string states, as
\begin{align}
B_{\text{non-grav}} = \frac{4}{\pi} \int_{\Lambda^2}^{\infty} d\mu \frac{\mathrm{Im}\, \mathcal{M}_\mathrm{w/o~Regge}(\mu + i\epsilon, t \to -0)}{(\mu - 2 m^2_\phi)^3} > 0.
\label{eq:non-grav_fin}
\end{align}
We will use Eq.~\eqref{eq:non-grav_fin} to estimate $B_{\text{non-grav}}$ in Sec.~\ref{sec:posi_bound}.

In Eq.~\eqref{eq:grav_positive}, $B_{\text{grav}}$ arises from the contribution involving a $t$-channel massless graviton (see panel (c) in Figure~\ref{fig:low_energy}).  
In many cases, $B_{\text{grav}}$ is negative~\cite{Cheung:2014vva,Cheung:2014ega,Andriolo:2018lvp,Alberte:2020jsk,Aoki:2021ckh,Noumi:2021uuv}.
Consequently, the gravitational positivity bound in Eq.~\eqref{eq:grav_positive} reads
\begin{align}
B_{\text{non-grav}} > |B_{\text{grav}}|.
\label{eq:grav_positive2}
\end{align}
%

Further discussions on the origin and validity of these bounds through string-loop effects in string-inspired models can be found in Refs.~\cite{Alberte:2020bdz,Alberte:2021dnj,Herrero-Valea:2022lfd,deRham:2022gfe,Hamada:2023cyt,Caron-Huot:2024tsk}.  
Related insights from the Weak Gravity Conjecture appear in Refs.~\cite{Hamada:2018dde,Andriolo:2018lvp,Chen:2019qvr,Arkani-Hamed:2021ajd,Noumi:2022ybv,Abe:2023anf}.

Thus, Eqs.~\eqref{eq:non-grav_fin} and \eqref{eq:grav_positive2} show that, if we assume only renormalizable interactions and no additional new physics above the scale $E > \Lambda$, 
then a violation of the positivity condition in Eq.~\eqref{eq:grav_positive2} at $\Lambda = E$ would imply that new physics at or above $E$ is required 
to raise the cutoff energy scale $\Lambda$ above $E$, 
since such new states must provide additional positive contributions to the left-hand side, $B_{\text{non-grav}}$. 
We adopt this assumption in Sec.~\ref{sec:posi_bound} to determine the energy scale up to which the renormalizable Higgs-portal DM remains valid.

Finally, let us comment on the conventional argument for positivity bounds.  
Dimension-8 EFT operators are included in $\tilde{a}_2(t \to -0)$ in Eq.~\eqref{eq:non-grav}, in addition to the Higgs-portal DM contributions.  
If new physics exists only at energy scales above $\Lambda$, then Eqs.~\eqref{eq:non-grav} and \eqref{eq:non-grav_fin} imply,
\begin{align}
\tilde{a}_2(t \to -0) = \frac{2}{\pi} \int_{\Lambda^2}^{\infty} d\mu \frac{\mathrm{Im}\, \mathcal{M}_\mathrm{w/o~Regge}(\mu + i\epsilon, t \to -0)}{(\mu - 2 m^2_\phi)^3} > 0.
\label{eq:dim8}
\end{align}
Here $\tilde{a}_2(t \to -0)$ corresponds to a combination of the Wilson coefficients of dimension-8 operators (for example, see \cite{Yamashita:2020gtt}).  
Eq.~\eqref{eq:dim8} shows that these dimension-8 contributions on the LHS arise from heavy states appearing at or above the energy scale $\Lambda$.
By unitarity of the new physics sector, these states provide
positive contributions to the RHS via $\mathrm{Im}\,\mathcal{M}$.
This is the standard argument for the original positivity bounds~\cite{Adams:2006sv}.

Before proceeding, we comment on recent developments regarding gravitational positivity bounds. Ref.~\cite{Hamada:2023cyt} proposed a one-loop Regge amplitude whose contribution may cancel the infrared-enhanced negative term appearing in gravitational positivity bounds. 
More recently, Caron-Huot and Tokuda demonstrated, in a string-inspired setup with Reggeized graviton exchange, 
that the large negative low-energy contributions enhanced by inverse powers of a light mass can indeed be cancelled by one-loop Regge contributions \cite{Caron-Huot:2024tsk}.
This observation suggests that deriving positivity constraints in gravitational theories requires a careful treatment of ultraviolet Regge physics. On the other hand, their analysis does not address a complete string embedding of the SM and leaves open the question of how positivity is realized in realistic particle-physics settings. Motivated by this situation, we adopt the working hypothesis that the positivity quantity $B(\Lambda)$ remains non-negative and investigate its phenomenological implications for DM physics. Our goal is not to establish the conjecture itself but to explore what classes of dark-sector scenarios would be suggested if such positivity holds. Throughout this work, we treat $B(\Lambda)\ge 0$ as a phenomenological working assumption rather than a rigorously established consequence of quantum gravity.

\subsection{Gravitational Positivity Bounds with $\phi \phi \to \phi \phi$}\label{sec:posi_bound}

\begin{figure}[!t]
\begin{center}
\includegraphics[width=0.65\textwidth,clip]{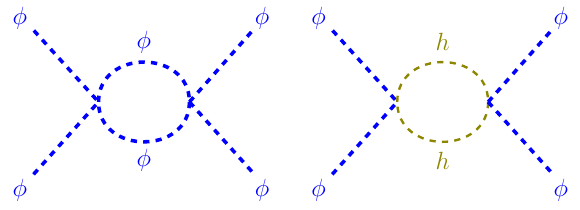}
\caption{Non-gravitational Feynman diagrams for $\phi \phi \to \phi \phi$}
\label{fig:diagram_non-grav}
\end{center}
\end{figure}
We now compute $B(\Lambda)$ for the $\phi \phi \to \phi \phi$ scattering process at the one-loop level, following the procedure described in Sec.~\ref{sec:posi_review}. 
The one-loop amplitudes are calculated using \texttt{FeynRules}~\cite{Christensen:2008py, Alloul:2013bka} and the \texttt{Mathematica} packages \texttt{FeynArts}~\cite{HAHN2001418}, \texttt{FeynCalc}~\cite{MERTIG1991345,Shtabovenko:2016sxi,Shtabovenko:2020gxv,Shtabovenko:2023idz}, and \texttt{Package-X}~\cite{Patel:2015tea}.

We first present the results for the non-gravitational contributions to the $\phi \phi \to \phi \phi$ scattering process. 
Figure~\ref{fig:diagram_non-grav} shows the leading non-gravitational one-loop Feynman diagrams for this process.

As illustrated there, only the Higgs-portal contributions are included, whereas possible contributions from new physics, i.e., heavy states, are not considered. 
(We do not include EFT operators in our calculation.)
Consequently, if our positivity bounds are violated at a certain energy scale $\Lambda = E$, 
additional new physics beyond the Higgs-portal interactions would be required to raise the cutoff energy scale $\Lambda$ above $E$, as discussed in Sec.~\ref{sec:posi_review}. 
If $(\Lambda=)E$ is sufficiently high, e.g., the grand unified theory scale or a typical string scale, this implies that 
the renormalizable Higgs-portal DM framework in Eq.~\eqref{eq:lag} (without any additional new physics) remains valid up to that high energy scale from the perspective of positivity.

The candy-like diagrams in Figure~\ref{fig:diagram_non-grav} contribute to $B_{\text{non-grav}}(\Lambda)$ with a 
suppression of order $1/\Lambda^4$, 
whereas other diagrams are further suppressed, e.g., by factors of $v^2/\Lambda^6$ or $v^4/\Lambda^8$. 
The dominant non-gravitational contribution, $B_{\text{non-grav}}(\Lambda)$, is therefore given by
\begin{align}
B^{\phi \phi \to \phi \phi}_{\text{non-grav}} &= \frac{\lambda_{h\phi}^2+\lambda_{\phi}^2}{16 \pi ^2 \Lambda ^4}.\label{eq:nongrav_B}
\end{align}

\begin{figure}[!t]
\begin{center}
\begin{minipage}[b]{0.25\linewidth}
\centering
\begin{tikzpicture}
\begin{feynman}
\vertex (a1) {\(\textcolor{blue}{\phi} \)};
\vertex at (1.2,0)(a2);
\vertex at (2.4,0)(a3);
\vertex at (3.6,0)(a4) {\(\textcolor{blue}{\phi} \)};
\vertex at (1.8,-1.0)(d);
\vertex at (0,-2.0) (c1){\(\textcolor{blue}{\phi} \)};
\vertex at (3.6,-2.0) (c3) {\(\textcolor{blue}{\phi} \)};
\vertex at (1.8,-2.0) (c2);
\diagram* {
(a1) -- [scalar, line width=0.6mm, color=blue] (a2) -- [scalar,  line width=0.6mm, color=blue, edge label=\(\textcolor{blue}{\phi} \)] (a3) -- [scalar, line width=0.6mm, color=blue] (a4),
(a2) -- [scalar, edge label'=\(h\),  line width=0.4mm, color=olive] (d) -- [scalar, edge label'=\(h\),  line width=0.4mm, color=olive] (a3),
(c1) -- [scalar, line width=0.6mm, color=blue] (c2)-- [scalar, line width=0.6mm, color=blue](c3),
(d) -- [graviton, edge label=\(G\), color=purple] (c2),
};
\end{feynman}
\end{tikzpicture}
\subcaption{}
\end{minipage}
\begin{minipage}[b]{0.25\linewidth}
\centering
\begin{tikzpicture}
\begin{feynman}
\vertex (a1) {\(\textcolor{blue}{\phi} \)};
\vertex at (1.2,0)(a2);
\vertex at (2.4,0)(a3);
\vertex at (3.6,0)(a4) {\(\textcolor{blue}{\phi} \)};
\vertex at (1.8,-1.0)(d);
\vertex at (0,-2.0) (c1){\(\textcolor{blue}{\phi} \)};
\vertex at (3.6,-2.0) (c3) {\(\textcolor{blue}{\phi} \)};
\vertex at (1.8,-2.0) (c2);
\diagram* {
(a1) -- [scalar, line width=0.6mm, color=blue] (a2) -- [scalar,  edge label=\(h\),  line width=0.4mm, color=olive] (a3) -- [scalar, line width=0.6mm, color=blue] (a4),
(a2) -- [scalar, line width=0.6mm, color=blue, edge label'=\(\textcolor{blue}{\phi} \)] (d) -- [scalar, line width=0.6mm, color=blue, edge label'=\(\textcolor{blue}{\phi} \)] (a3),
(c1) -- [scalar, line width=0.6mm, color=blue] (c2)-- [scalar, line width=0.6mm, color=blue](c3),
(d) -- [graviton, edge label=\(G\), color=purple] (c2),
};
\end{feynman}
\end{tikzpicture}
\subcaption{}
\end{minipage}\\
\begin{minipage}[b]{0.25\linewidth}
\centering
\begin{tikzpicture}
\begin{feynman}
\vertex at (0.0,-2.0)(a1) {\(\textcolor{blue}{\phi} \)};
\vertex at (1.2,-2.0)(a2);
\vertex at (2.4,-2.0)(a3);
\vertex at (3.6,-2.0)(a4) {\(\textcolor{blue}{\phi} \)};
\vertex at (1.8,-1.0)(d);
\vertex at (0,0.0) (c1){\(\textcolor{blue}{\phi} \)};
\vertex at (3.6,0.0) (c3) {\(\textcolor{blue}{\phi} \)};
\vertex at (1.8,0.0) (c2);
\diagram* {
(a1) -- [scalar, line width=0.6mm, color=blue] (a2) -- [scalar,  line width=0.6mm, color=blue, edge label'=\(\textcolor{blue}{\phi} \)] (a3) -- [scalar, line width=0.6mm, color=blue] (a4),
(a2) -- [scalar, edge label=\(h\),  line width=0.4mm, color=olive] (d) -- [scalar, edge label=\(h\),  line width=0.4mm, color=olive] (a3),
(c1) -- [scalar, line width=0.6mm, color=blue] (c2)-- [scalar, line width=0.6mm, color=blue](c3),
(d) -- [graviton, edge label'=\(G\), color=purple] (c2),
};
\end{feynman}
\end{tikzpicture}
\subcaption{}
\end{minipage}
\begin{minipage}[b]{0.25\linewidth}
\centering
\begin{tikzpicture}
\begin{feynman}
\vertex at (0,-2.0)(a1) {\(\textcolor{blue}{\phi} \)};
\vertex at (1.2,-2.0)(a2);
\vertex at (2.4,-2.0)(a3);
\vertex at (3.6,-2.0)(a4) {\(\textcolor{blue}{\phi} \)};
\vertex at (1.8,-1.0)(d);
\vertex at (0,0) (c1){\(\textcolor{blue}{\phi} \)};
\vertex at (3.6,0) (c3) {\(\textcolor{blue}{\phi} \)};
\vertex at (1.8,0) (c2);
\diagram* {
(a1) -- [scalar, line width=0.6mm, color=blue] (a2) -- [scalar,  edge label'=\(h\),  line width=0.4mm, color=olive] (a3) -- [scalar, line width=0.6mm, color=blue] (a4),
(a2) -- [scalar, line width=0.6mm, color=blue, edge label=\(\textcolor{blue}{\phi} \)] (d) -- [scalar, line width=0.6mm, color=blue, edge label=\(\textcolor{blue}{\phi} \)] (a3),
(c1) -- [scalar, line width=0.6mm, color=blue] (c2)-- [scalar, line width=0.6mm, color=blue](c3),
(d) -- [graviton, edge label'=\(G\), color=purple] (c2),
};
\end{feynman}
\end{tikzpicture}
\subcaption{}
\end{minipage}
\caption{$t$-channel graviton exchange Feynman diagrams for $\phi \phi \to \phi \phi$}
\label{fig:diagram_grav}
\end{center}
\end{figure}
Next, we present the gravitational contribution, $B_{\text{grav}}(\Lambda)$, for the $\phi \phi \to \phi \phi$ process. 
Figure~\ref{fig:diagram_grav} shows the leading one-loop gravitational Feynman diagrams in the $t$-channel. 
The dominant gravitational contribution is given by
\begin{align}
B^{\text{1-loop}}_{\text{grav}} &= B^{\text{1-loop}}_{\text{grav(a)}} + B^{\text{1-loop}}_{\text{grav(b)}} + B^{\text{1-loop}}_{\text{grav(c)}} + B^{\text{1-loop}}_{\text{grav(d)}} = -\frac{\lambda^2_{h\phi} v^2}{24 \pi^2\overline{M}^2_{\text{pl}} m^4_h}f(m_{\phi}/m_h), \label{eq:grav_B_1-loop} \\
f(x) &= \frac{1}{x^6(1 - 4x^2)^2}\biggl\{28 x^6 -15 x^4 +2 x^2-(1-4 x^2)^2(x^2-1) \ln(x^2)\nonumber\\
&\, \, \, \, \, \, - 2\sqrt{1-4 x^2}(2x^6-12 x^4+7 x^2-1)\ln\left[\frac{1}{2x}\left(\sqrt{1-4x^2}+1\right)\right]\biggr\},
\end{align}
where $\overline{M}_{\text{pl}} = M_{\text{pl}}/\sqrt{8\pi} \sim 2.4 \times 10^{18}~\mathrm{GeV}$ is the reduced Planck mass, and $m_h$ is the Higgs boson mass. 
Explicit expressions for each contribution, i.e., $B^{\text{1-loop}}_{\text{grav(a/b/c/d)}}$, are provided in Appendix~\ref{sec:appendix_B}.
\begin{figure}[t!]\center
\includegraphics[width=0.495\textwidth,clip]{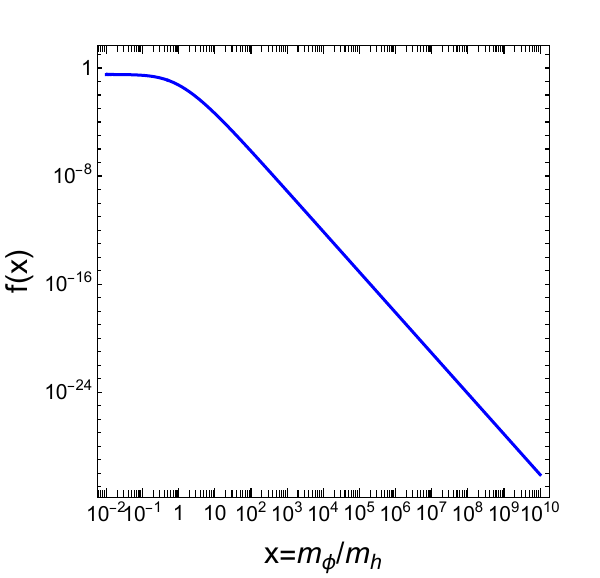}
\includegraphics[width=0.495\textwidth,clip]{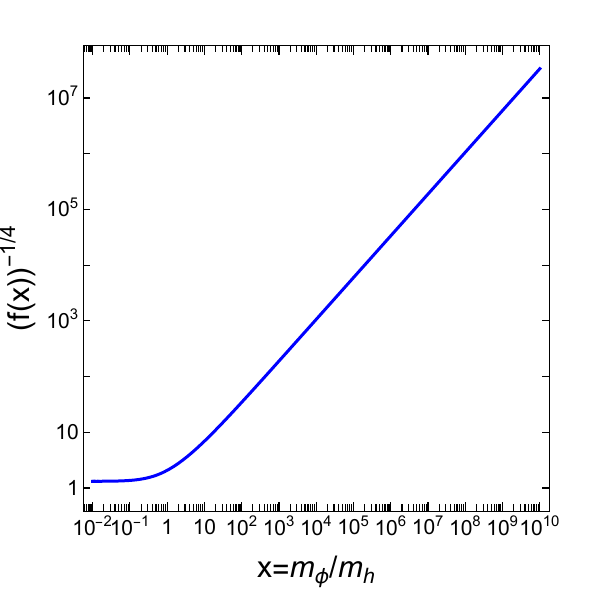}
\caption{Plots of $f(x)$ (left) and $f^{-1/4}(x)$ (right). The function $f(x)$ is real and positive in the plotted range.}
\label{fig:fx}
\vspace{5mm}
\end{figure}
The function $f(x)$ is displayed in the left panel of Figure~\ref{fig:fx}. 
As can be seen from the plot, this function is real and positive. 
This implies that the gravitational contribution in Eq.~\eqref{eq:grav_B_1-loop} is negative.

\begin{figure}[t]
\centering
\begin{tikzpicture}
\begin{feynman}
\vertex (a1) {\(\textcolor{blue}{\phi} \)};
\vertex at (1.2,0)(a2);
\vertex at (2.4,0)(a3);
\vertex at (3.6,0)(a4) {\(\textcolor{blue}{\phi} \)};
\vertex at (1.8,-0.6)(d);
\vertex at (0,-1.6) (c1){\(\textcolor{blue}{\phi} \)};
\vertex at (1.8,-1.6) (c2);
\vertex at (3.6,-1.6) (c3) {\(\textcolor{blue}{\phi} \)};
\diagram* {
(a1) -- [scalar, line width=0.6mm, color=blue] (a2) -- [scalar, line width=0.6mm, color=blue] (a4),
(a2) -- [scalar,  line width=0.6mm, color=blue, half left](a3) -- [scalar, line width=0.6mm, color=blue, quarter left](d) -- [scalar, line width=0.6mm, color=blue, quarter left](a2), 
(c1) -- [scalar, line width=0.6mm, color=blue] (c2)-- [scalar, line width=0.6mm, color=blue](c3),
(d) -- [graviton, edge label=\(G\), color=purple] (c2), 
};
\end{feynman}
\end{tikzpicture}
\caption{Two-loop contribution to the forward scattering process $\phi\phi\to\phi\phi$, involving only the real scalar DM field $\phi$, corresponding to the leading contribution from the four-point self-coupling in the gravitational positivity bound.}
\label{fig:phi_2loop}
\end{figure}
From Eq.~\eqref{eq:grav_B_1-loop}, we can see that only the Higgs-portal coupling contributes to the gravitational correction $B^{\text{1-loop}}_{\text{grav}}$.
The contribution from the four-point self-coupling first appears at the two-loop level and provides the leading contribution, as depicted in Figure~\ref{fig:phi_2loop} and given by~\cite{Noumi:2021uuv}
\begin{align}
B^{\text{2-loop}}_{\text{grav}} = - \frac{10-\pi^2}{4608\pi^4}\frac{\lambda^2_{\phi}}{\overline{M}^2_{\text{pl}}m^2_\phi}.
\label{eq:grav_B_2-loop}
\end{align}
Then, 
\begin{align}
B^{\phi \phi \to \phi \phi}_{\text{grav}} &= B^{\text{1-loop}}_{\text{grav}} + B^{\text{2-loop}}_{\text{grav}}\\
&= -\frac{\lambda^2_{h\phi} v^2}{24 \pi^2\overline{M}^2_{\text{pl}} m^4_h}f(m_{\phi}/m_h) - \frac{10-\pi^2}{4608\pi^4}\frac{\lambda^2_{\phi}}{\overline{M}^2_{\text{pl}}m^2_\phi}.\label{eq:grav_B}
\end{align}

Combining Eqs.~\eqref{eq:nongrav_B} and \eqref{eq:grav_B}, the gravitational positivity bound for the $\phi \phi \to \phi \phi$ process is given by
\begin{align}
B^{\phi \phi \to \phi \phi}(\Lambda)
&= B^{\phi \phi \to \phi \phi}_{\text{non-grav}}
 + B^{\phi \phi \to \phi \phi}_{\text{grav}}
 \geq 0, \notag\\
\frac{\lambda_{h\phi}^2 + \lambda_{\phi}^2}
{16\pi^2 \Lambda^4}
&\geq
\frac{\lambda^2_{h\phi} v^2}
{24\pi^2 \overline{M}^2_{\text{pl}} m^4_h}
f(m_{\phi}/m_h)
+ \frac{10-\pi^2}{4608\pi^4}
\frac{\lambda^2_{\phi}}
{\overline{M}^2_{\text{pl}} m^2_\phi},\\
\lambda_{h\phi}^2 + \lambda_{\phi}^2
&\geq
\frac{2 \Lambda^4}
{3 \overline{M}^2_{\text{pl}}}
\left(
\frac{\lambda^2_{h\phi} v^2}{m^4_h}
f(m_\phi/m_h)
+ \frac{10-\pi^2}{192\pi^2}
\frac{\lambda^2_{\phi}}{m^2_\phi}
\right),
\label{eq:cond0}
\end{align}
which explicitly leads to the upper bound
\begin{align}
\Lambda \le
\left[
\frac{3\,\overline{M}_{\mathrm{pl}}^{2} m_h^{4} m_\phi^{2}}{2}
\,
\frac{
192\pi^2\left(\lambda_{h\phi}^2 + \lambda_\phi^{2}\right)
}{
192\pi^2 \lambda_{h\phi}^2 v^2 m_\phi^{2}
f\!\left(\frac{m_\phi}{m_h}\right)
+
(10-\pi^2)\lambda_\phi^{2} m_h^{4}
}
\right]^{1/4}.
\label{eq:cond}
\end{align}

From Eq.~\eqref{eq:cond}, when 
$\lambda_{h\phi} \neq 0$ and $\lambda_\phi = 0$, 
i.e., in the presence of the Higgs-portal interaction without scalar DM self-interactions, we obtain
\begin{align}
\Lambda &\leq \left(\frac{3}{2}\right)^{1/4} \left(\frac{\overline{M}_{\text{pl}}}{v}\right)^{1/2} f^{-1/4}(m_\phi/m_h)m_h, \label{eq:cond2} \\
\Lambda &\lesssim 10^{10}~\mathrm{GeV} \, \text{ for }\, m_\phi < m_h. 
\label{eq:cond2-2}
\end{align}
Equation~\eqref{eq:cond2-2} indicates that new physics is expected to appear below an energy scale of approximately $10^{10}$~GeV when the Higgs-portal scalar DM has no self-interaction and a mass satisfying $m_\phi < m_h$.
Here we assume that $f^{-1/4}(m_\phi/m_h)$ is of order $\mathcal{O}(1)$.  
Note that $f(x) \to 1/3$ as $x \to 0$, which corresponds to $f^{-1/4}(x) \simeq 1.3$.
Therefore, this assumption is valid in the light DM regime.
The function $f^{-1/4}(x)$, shown in the right panel of Figure~\ref{fig:fx}, indeed takes values of order $\mathcal{O}(1)$ for $m_\phi < m_h$, supporting the estimate in Eq.~\eqref{eq:cond2-2}.
As the figure suggests, a heavier DM mass allows the cutoff scale $\Lambda$ to approach the GUT scale $\Lambda_{\rm GUT}$, i.e., $10^{16}$~GeV.
Indeed, this occurs once the DM mass exceeds approximately $10^{10}$~GeV.
%
\begin{figure}[t!]\center
\includegraphics[width=0.495\textwidth,clip]{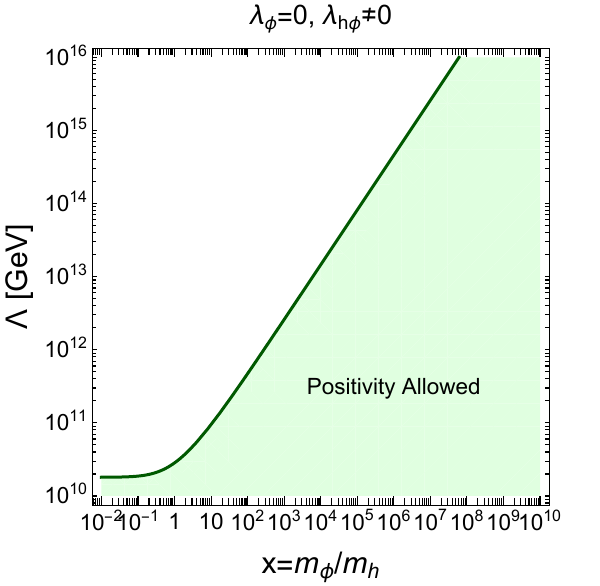}
\includegraphics[width=0.495\textwidth,clip]{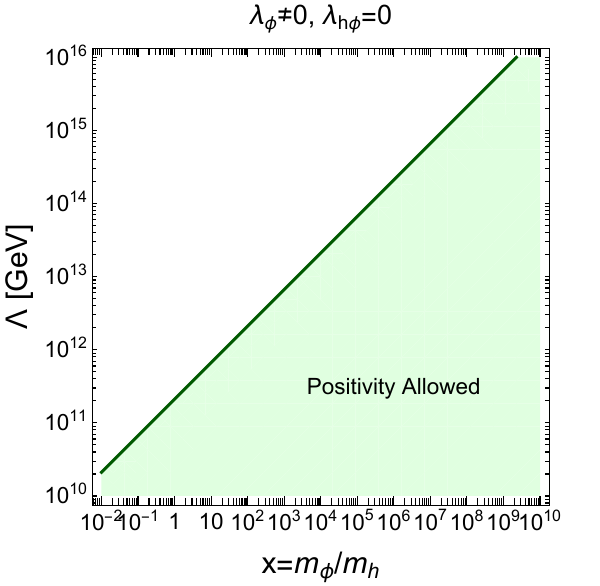}
\caption{
Parameter space of $m_\phi/m_h$ versus $\Lambda$ satisfying the gravitational positivity bound from the $\phi\phi \to \phi\phi$ process.
The left (right) panel corresponds to $\lambda_\phi = 0$ ($\lambda_{h\phi} = 0$).
The green region indicates the parameter space where the bound in Eq.~\eqref{eq:cond} is satisfied.
}
\label{fig:cutoff-ratio}
\end{figure}

\begin{figure}[t!]\center
\includegraphics[width=0.495\textwidth,clip]{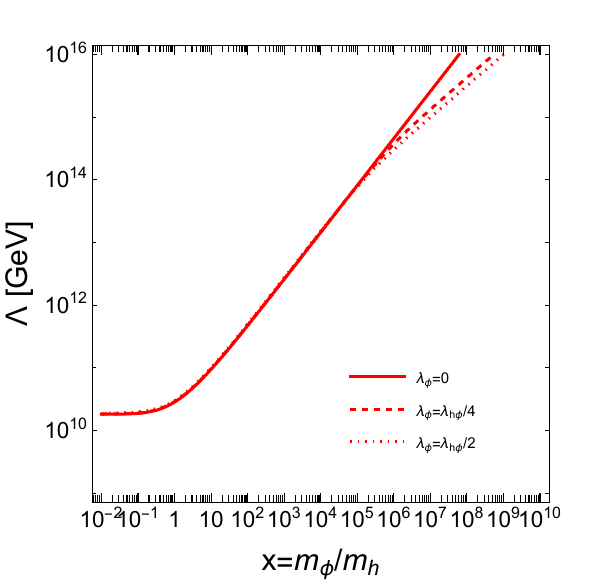}
\includegraphics[width=0.495\textwidth,clip]{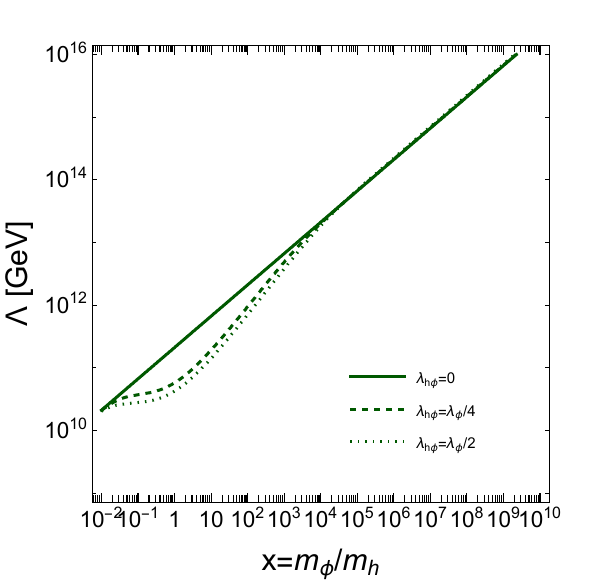}
\caption{
Parameter space of $m_\phi/m_h$ versus $\Lambda$ satisfying the gravitational positivity bound from the $\phi\phi \to \phi\phi$ process.
In the left panel, the curves correspond to $\lambda_\phi = 0$ (red solid), $\lambda_\phi = \lambda_{h\phi}/4$ (red dashed), and $\lambda_\phi = \lambda_{h\phi}/2$ (red dotted).
In the right panel, the curves correspond to $\lambda_{h\phi} = 0$ (dark green solid), $\lambda_{h\phi} = \lambda_\phi/4$ (dark green dashed), and $\lambda_{h\phi} = \lambda_\phi/2$ (dark green dotted). For each curve, the region below the curve satisfies the bound in Eq.~\eqref{eq:cond}.
}
\label{fig:posi_general}
\end{figure}

More generally, writing $r\equiv\lambda_\phi/\lambda_{h\phi}$ and
$x\equiv m_\phi/m_h$, Eq.~\eqref{eq:cond0} at $\Lambda=\Lambda_{\rm GUT}$ becomes an
implicit equation for the minimal DM mass as a function of $r$,
\begin{equation}
1+r^2 = \frac{2\Lambda_{\rm GUT}^4}{3\overline M_{\rm pl}^2}
\left[\frac{v^2}{m_h^4}f(x) + \frac{10-\pi^2}{192\pi^2}\frac{r^2}{m_\phi^2}\right],
\label{eq:mphi_of_r}
\end{equation}
which reduces to Eq.~\eqref{eq:cond2-2} for $r=0$.
In the opposite limit,
$r\to\infty$, Eq.~\eqref{eq:mphi_of_r} reduces to the closed form
\begin{equation}
m_\phi \gtrsim \sqrt{\frac{2\Lambda_{\rm GUT}^4}{3\overline M_{\rm pl}^2}\cdot
\frac{10-\pi^2}{192\pi^2}} \simeq 2.8\times10^{11}\,{\rm GeV}
\label{eq:mphi_r_infty}
\end{equation}
consistent with the right panel of Figure~\ref{fig:cutoff-ratio}. Solving
Eq.~\eqref{eq:mphi_of_r} numerically for intermediate values of $r$
gives the minimal masses listed in Table~\ref{tab:n}, interpolating monotonically
between these two limits and consistent with the curves shown in
Figure~\ref{fig:posi_general}.

\begin{table}[h]
\centering
\begin{tabular}{c|cccccccc}
$r=\lambda_\phi/\lambda_{h\phi}$ & 0 & 0.25 & 0.5 & 1 & 2 & 4 & 10 & $\infty$ \\\hline
$m_\phi^{\rm min}\,[10^{10}\,{\rm GeV}]$ & 0.76 & 6.8 & 12.6 & 20.0 & 25.2 & 27.4 & 28.1 & 28.2
\end{tabular}
\caption{Minimal DM mass required to reach $\Lambda_{\rm GUT}=10^{16}\,{\rm GeV}$,
as a function of $r=\lambda_\phi/\lambda_{h\phi}$, obtained by solving
Eq.~\eqref{eq:mphi_of_r}.}\label{tab:n}
\end{table}


%
The parameter space of $m_\phi/m_h$ versus $\Lambda$ is illustrated in Figure~\ref{fig:cutoff-ratio}.
The left panel corresponds to the pure Higgs-portal case, $\lambda_\phi = 0$, whereas the right panel represents the pure DM four-point coupling case, $\lambda_{h\phi} = 0$.
In both panels, the green shaded region satisfies the gravitational positivity bound given in Eq.~\eqref{eq:cond}.
As anticipated from Eq.~\eqref{eq:cond}, a comparison between the cases $\lambda_\phi = 0$ (left) and $\lambda_{h\phi} = 0$ (right) in Figure~\ref{fig:cutoff-ratio} shows that achieving a cutoff scale at the GUT scale, $\Lambda_{\text{GUT}}$, requires the DM mass to be approximately $10^{10}$~GeV (left) and $10^{11}$~GeV (right), or larger.

Next, we summarize the more general case in which both the Higgs-portal coupling and the DM four-point self-coupling are present.
\begin{figure}[t!]\center
\includegraphics[width=0.495\textwidth,clip]{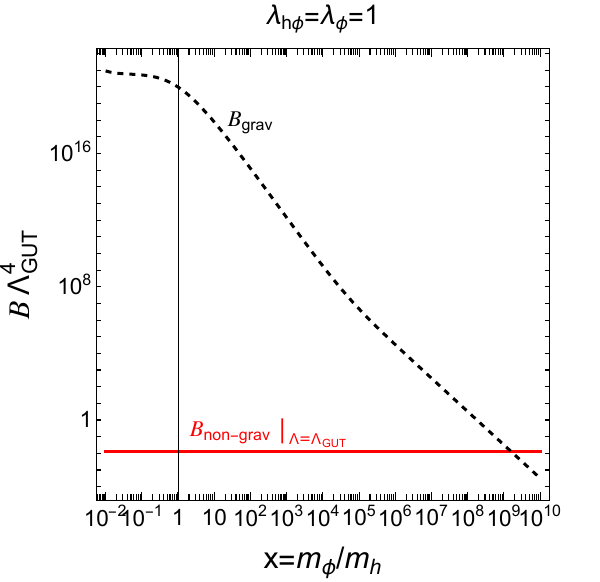}
\includegraphics[width=0.495\textwidth,clip]{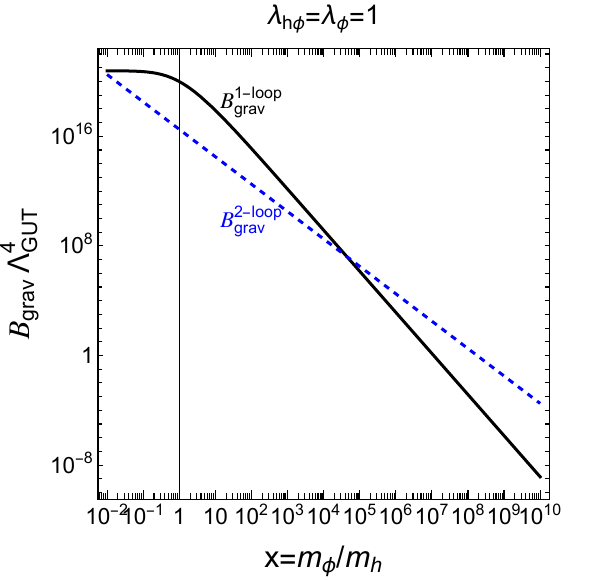}
\caption{
Parameter space of $m_\phi/m_h$ versus $B\Lambda_{\text{GUT}}^4$.
The left panel compares $B_{\text{grav}}$ (black dashed line) and $B_{\text{non-grav}}$ evaluated with a GUT-scale cutoff (red solid line).
The right panel shows the one-loop (black solid line) and two-loop (blue dashed line) contributions to $B_{\text{grav}}$.
}
\label{fig:B}
\end{figure}

Figure~\ref{fig:B} illustrates the comparison between the gravitational contribution $B_{\text{grav}}$ (black dashed line) and the non-gravitational contribution $B_{\text{non-grav}}$ evaluated with a GUT-scale cutoff (red solid line) as functions of $m_\phi/m_h$ in the left panel.
The right panel displays the one-loop (black solid line) and two-loop (blue dashed line) contributions to $B_{\text{grav}}$.
In both panels, we take $\lambda_{h\phi} = \lambda_\phi = 1$.
Changing the common value of the couplings from unity rescales the vertical axis but does not affect the relative contributions to $B$.
From the left panel, we find that the gravitational positivity bound is satisfied for DM masses of order $10^{11}\,\mathrm{GeV}$ or larger when the two couplings are equal.
From the right panel, we observe that the two-loop contribution to $B_{\text{grav}}$ becomes dominant for DM masses around $10^7\,\mathrm{GeV}$.
This behavior is consistent with the case in which only the DM self-coupling is present.

Finally, we consider the case in which the two couplings are varied away from the same value.
Figure~\ref{fig:posi_general} shows the parameter space of $m_\phi/m_h$ versus $\Lambda$ satisfying the gravitational positivity bound from the $\phi\phi \to \phi\phi$ process. For each curve, the parameter region below the curve satisfies the bound in Eq.~\eqref{eq:cond}. The left panel demonstrates how increasing the DM self-coupling shifts the minimum DM mass required to realize a GUT-scale cutoff from approximately $10^{10}\,\mathrm{GeV}$ to $10^{11}\,\mathrm{GeV}$. The right panel shows that increasing the Higgs-portal coupling affects the minimum cutoff scale primarily for DM masses near the Higgs boson mass, but does not lead to a dramatic change.

We therefore conclude that, in the presence of both interactions, achieving a cutoff scale at the GUT scale generally requires a DM mass of order $10^{10}$--$10^{11}\,\mathrm{GeV}$ (or above), with the larger values favored when the four-point self-coupling plays a significant role.

Let us comment on the possible additional contribution on the RHS of Eq.~\eqref{eq:grav_positive}, which could potentially lead to a violation of the positivity bound.
When setting the GUT scale as the cutoff scale, $\Lambda = \Lambda_{\text{GUT}}$, the contribution of order
\(
\mathcal{O}\!\left(1/(\Lambda_{\mathrm{QG}}^{2} \overline{M}^2_{\text{pl}})\right)
= \mathcal{O}(\alpha'/\overline{M}^2_{\text{pl}})
\)
on the RHS of Eq.~\eqref{eq:grav_positive} can be safely neglected.
Here we assume a typical value of $\Lambda_{\mathrm{QG}} = \alpha'^{-1/2} \sim 10^{16}\,\mathrm{GeV}$, which represents the scale of Reggeization~\cite{Hamada:2018dde,Tokuda:2020mlf,Aoki:2021ckh,Noumi:2021uuv}.
In this case, the applicability condition
\[
\min(\Lambda, \Lambda_{*}) < \sqrt{\overline{M}_{\text{pl}}\,\Lambda_{\mathrm{QG}}}
\]
is satisfied,
where $\Lambda_{*} = B_{\text{grav}}^{-1/4}$~\cite{Noumi:2021uuv}.

\section{Phenomenological Implications for Heavy Dark Matter}\label{sec:dm_pheno}
Based on the results presented in Sec.~\ref{sec:posi_bound},
we discuss the phenomenological implications of the gravitational
positivity bounds for the Higgs-portal real scalar DM model
in the heavy-mass regime.
If the new physics scale $\Lambda$ is identified with the GUT scale
or a typical string scale, $\Lambda \sim 10^{16}\,\mathrm{GeV}$,
the bound implies a DM mass of order
$10^{10}$--$10^{11}\,\mathrm{GeV}$ (or above).

The Higgs-portal coupling
is constrained from the perspective of the DM relic abundance.
There may exist additional scattering processes that lead to further
gravitational positivity bounds, for example by constraining the
couplings; we leave such analyses for future work.

Since the DM mass is extremely large, the conventional
weakly interacting massive particle (WIMP) scenario
cannot reproduce the observed relic abundance due to the
perturbative unitarity bound on the Higgs-portal coupling.

We therefore consider the feebly interacting massive particle (FIMP)
scenario, in which DM never thermalizes with the
SM plasma, following Ref.~\cite{Aoki:2022dzd}.
Here we focus on DM production after reheating.
DM production during the reheating era depends on the
inflationary dynamics and sets the initial condition for subsequent
production.

When $T_{\mathrm{reh}} > m_\phi$ and the DM is heavy,
gravitational interactions can also contribute to DM
production in addition to the Higgs-portal interaction.
We therefore include this contribution in our analysis.

Gravitational interactions are described by
\begin{align}
\mathcal{L}_{\text{grav}}
=
\frac{1}{\overline{M}_{\text{pl}}}
h^{\mu\nu}
\left(
T^{\mathrm{SM}}_{\mu\nu}
+
T^{h}_{\mu\nu}
+
T^{\phi}_{\mu\nu}
\right),
\label{int_h}
\end{align}
where $h^{\mu\nu}$ denotes the graviton field.
The energy--momentum tensor for a real scalar field $S$ is given by
\begin{align}
T^{S}_{\mu\nu}
= \partial_{\mu} S \, \partial_{\nu} S
-
\frac{1}{2}
\eta_{\mu\nu}
\eta^{\rho\sigma} \partial_{\rho} S \, \partial_{\sigma} S
-
\frac{1}{2}
\eta_{\mu\nu}
m_{S}^{2} S^{2},
\qquad
(S = h,\, \phi).
\end{align}

Based on the interaction in Eq.~\eqref{int_h}, the scalar DM particle $\phi$ can be produced from the SM plasma.
The scattering amplitude for the process $h + h \to \phi + \phi$ via graviton exchange is given by~\cite{Aoki:2022dzd}
\begin{align}
\mathcal{M}^{G}_{h+h \to \phi+\phi}
=
-
\frac{1}{\overline{M}_{\text{pl}}^{2}}
\frac{(t - m_{\phi}^{2})(s + t - m_{\phi}^{2})}{s},
\end{align}
where $s$ and $t$ are the Mandelstam variables.
Including the Higgs-portal interaction, the total squared amplitude for $h + h \to \phi + \phi$ becomes
\begin{align}
\left|
\mathcal{M}^{\mathrm{total}}_{h+h \to \phi+\phi}
\right|^{2}
=
\left(
\lambda_{h\phi}
+
\frac{(t - m_{\phi}^{2})(s + t - m_{\phi}^{2})}
{s\,\overline{M}_{\text{pl}}^{2}}
\right)^{2}.
\label{M_hGX}
\end{align}

Similarly, the contributions from other SM particles through graviton exchange are given by~\cite{Aoki:2022dzd}
\begin{align}
\left|
\mathcal{M}^{G}_{f+f \to \phi+\phi}
\right|^{2}
&=
-
\frac{1}{2 \overline{M}_{\text{pl}}^{4} s^{2}}
\left(
s + 2t - 2 m_{\phi}^{2}
\right)^{2}
\left[
\left(
t - m_{\phi}^{2}
\right)^{2}
+ s t
\right],
\label{M_fGX}
\\[4pt]
\left|
\mathcal{M}^{G}_{V+V \to \phi+\phi}
\right|^{2}
&=
\frac{2}{\overline{M}_{\text{pl}}^{4} s^{2}}
\left(
m_{\phi}^{4}
-
2 m_{\phi}^{2} t
+
t (s + t)
\right)^{2},
\label{M_VGX}
\end{align}
where $f$ and $V$ denote SM fermions and SM vector bosons, respectively.

When DM is decoupled from the SM plasma, its number density $n_\phi$ evolves according to the Boltzmann equation with a production reaction rate~\cite{Hall:2009bx}
\begin{align}
\dot{n}_\phi + 3 H n_\phi = R(T),
\label{DMeq}
\end{align}
where $R(T)$ denotes the reaction rate for thermal scattering processes.

For the thermal production process
$i_1(p_1) + i_2(p_2) \to \phi(p_3) + \phi(p_4)$,
the reaction rate can be written in terms of the squared amplitude
$\overline{|\mathcal{M}_{i_1+i_2 \to \phi+\phi}|^2}$ as~\cite{Hall:2009bx,Edsjo:1997bg}
\begin{align}
R(T)
=
\frac{T}{2^{11}\pi^{6}}
\int_{4 m_\phi^{2}}^{\infty} \! ds \int d\Omega \,
K_{1}\!\left(\frac{\sqrt{s}}{T}\right)
\sqrt{s - 4 m_\phi^{2}}\,
\overline{|\mathcal{M}_{i_1+i_2 \to \phi+\phi}|^{2}},
\label{R_T}
\end{align}
where $i_{1,2}$ collectively denote the SM radiation.
The solid angle element is defined as
$d\Omega \equiv 2\pi\, d\cos\theta_{13}$, with $\theta_{13}$ being the angle between the momenta $\mathbf{p}_1$ and $\mathbf{p}_3$ in the center-of-mass frame,
and $K_1(z)$ is the modified Bessel function of the second kind.
The overline indicates that appropriate symmetry factors for the initial and final states are included.

After reheating, using $T \propto a^{-1}$ (and hence $\dot{T} = - H T$) together with
\begin{align}
H
=
\sqrt{\frac{g_{\mathrm{reh}} \pi^{2}}{90}}
\frac{T^{2}}{\overline{M}_{\text{pl}}},
\end{align}
where $g_{\mathrm{reh}}=106.75$ denotes the effective number of relativistic degrees of freedom during reheating,
the Boltzmann equation~\eqref{DMeq} can be rewritten in terms of the DM yield
$Y \equiv n_\phi / T^{3}$ as
\begin{align}
\frac{dY}{dT}
=
- \frac{1}{H T^{4}} R(T)
=
- \sqrt{\frac{90}{\pi^{2} g_{\mathrm{reh}}}}
\frac{\overline{M}_{\text{pl}}}{T^{6}} R(T).
\label{DMeq2}
\end{align}

The total reaction rate can be expressed as
\begin{align}
R_{\rm{total}}(T)=4R_h(T)+45R_f(T)+12R_V(T),  
\label{eq:totR}
\end{align}
where $R_h$, $R_f$, and $R_V$ are the reaction rates defined in Eq.~\eqref{R_T}, corresponding to the matrix elements given in Eqs.~\eqref{M_hGX}, \eqref{M_fGX}, and \eqref{M_VGX}, respectively.
The numerical coefficients account for the relevant SM degrees of freedom.
Here we consider the Higgs-unbroken phase.

From Ref.~\cite{Aoki:2022dzd}, the combined contribution from SM fermions and vector bosons is given by
\begin{align}
45R_f(T)+12R_V(T) = 
\frac{69 m_{\phi}^7 T}{512 \pi ^{9/2} M^4_{\text{pl}}}G_{1,3}^{3,0}\left.\left(  \begin{array}{c}
0 \\
-\frac{7}{2},-\frac{1}{2}, \frac{1}{2}
\end{array}
\right| \frac{m_{\phi}^{2} }{T^{2}}\right),
\label{eq:RfV}
\end{align}
where $G_{p, q}^{m, n}\left.\left(\begin{array}{c}
a_{1}, \ldots, a_{p} \\
b_{1}, \ldots, b_{q}
\end{array} \right| z\right)$ denotes the Meijer $G$-function. 

We obtain the Higgs contribution
\begin{align}
4R_h(T)
&=
\frac{\lambda_{h\phi}^{2}\, m_{\phi}^{2} T^{2}}
{128\,\pi^{5}}
K_{1}^{2}\!\left(\frac{m_{\phi}}{T}\right)
\nonumber\\[6pt]
&\quad
+
\frac{m_{\phi}^{2} T}{1920\,\pi^{5} \overline{M}_{\text{pl}}^{4}}
\Biggl[
4\sqrt{\pi}\, m_{\phi}^{5}\,
G_{1,3}^{3,0}\!\left(
\left.
\begin{array}{c}
-2 \\
-\frac{7}{2},\, -\frac{1}{2},\, \frac{1}{2}
\end{array}
\right|
\frac{m_{\phi}^{2}}{T^{2}}
\right)
\nonumber\\
&\qquad\qquad
+\,2\sqrt{\pi}\, m_{\phi}^{5}\,
G_{1,3}^{3,0}\!\left(
\left.
\begin{array}{c}
-1 \\
-\frac{5}{2},\, -\frac{1}{2},\, \frac{1}{2}
\end{array}
\right|
\frac{m_{\phi}^{2}}{T^{2}}
\right)
+\,3 m_{\phi}^{4} T\,
K_{1}^{2}\!\left(\frac{m_{\phi}}{T}\right)
\Biggr]
\nonumber\\[6pt]
&\quad
+
\frac{\lambda_{h\phi}\, m_{\phi}^{2} T}
{1920\,\pi^{5} \overline{M}_{\text{pl}}^{4}}
\Biggl[
-10\sqrt{\pi}\, \overline{M}_{\text{pl}}^{2} m_{\phi}^{3}\,
G_{1,3}^{3,0}\!\left(
\left.
\begin{array}{c}
-1 \\
-\frac{5}{2},\, -\frac{1}{2},\, \frac{1}{2}
\end{array}
\right|
\frac{m_{\phi}^{2}}{T^{2}}
\right)
\nonumber\\
&\qquad\qquad
-10 \overline{M}_{\text{pl}}^{2} m_{\phi}^{2} T\,
K_{1}^{2}\!\left(\frac{m_{\phi}}{T}\right)
\Biggr].
\label{eq:Rh}
\end{align}
The first term in Eq.~\eqref{eq:Rh} originates from the Higgs-portal
interaction contribution to $hh \to \phi\phi$.
The second square-bracketed term arises from purely gravitational
interactions, and the third square-bracketed term corresponds to
their interference.

Combining Eqs.~\eqref{eq:RfV} and \eqref{eq:Rh},
the total reaction rate in Eq.~\eqref{eq:totR} can be written as
\begin{align}
R_{\text{tot}}(T)
&=
R_{\text{portal}}(T)
+
R_{\text{grav}}(T)
+
R_{\text{int}}(T),
\label{eq:totR2}
\\[6pt]
R_{\text{portal}}(T)
&=
\frac{\lambda_{h\phi}^{2}\, m_{\phi}^{2} T^{2}}
{128\,\pi^{5}}\,
K_{1}^{2}\!\left(\frac{m_{\phi}}{T}\right),
\\[6pt]
R_{\text{grav}}(T)
&=
\frac{m_{\phi}^{2} T}
{1920\,\pi^{5} \overline{M}_{\text{pl}}^{4}}
\Biggl[
4\sqrt{\pi}\, m_{\phi}^{5}\,
G_{1,3}^{3,0}\!\left(
\left.
\begin{array}{c}
-2\\
-\frac72,\,-\frac12,\,\frac12
\end{array}
\right|
\frac{m_{\phi}^{2}}{T^{2}}
\right)
\nonumber\\
&\qquad
+ 2\sqrt{\pi}\, m_{\phi}^{5}\,
G_{1,3}^{3,0}\!\left(
\left.
\begin{array}{c}
-1\\
-\frac52,\,-\frac12,\,\frac12
\end{array}
\right|
\frac{m_{\phi}^{2}}{T^{2}}
\right)
+ 3 m_{\phi}^{4} T\,
K_{1}^{2}\!\left(\frac{m_{\phi}}{T}\right)
\Biggr]
\nonumber\\
&\quad
+
\frac{69\, m_{\phi}^{7} T}
{512\,\pi^{9/2} \overline{M}_{\text{pl}}^{4}}
G_{1,3}^{3,0}\!\left(
\left.
\begin{array}{c}
0\\
-\frac72,\,-\frac12,\,\frac12
\end{array}
\right|
\frac{m_{\phi}^{2}}{T^{2}}
\right),
\\[6pt]
R_{\text{int}}(T)
&=
\frac{\lambda_{h\phi}\, m_{\phi}^{2} T}
{1920\,\pi^{5} \overline{M}_{\text{pl}}^{4}}
\Biggl[
-10\sqrt{\pi}\, \overline{M}_{\text{pl}}^{2} m_{\phi}^{3}\,
G_{1,3}^{3,0}\!\left(
\left.
\begin{array}{c}
-1\\
-\frac52,\,-\frac12,\,\frac12
\end{array}
\right|
\frac{m_{\phi}^{2}}{T^{2}}
\right)
\nonumber\\
&\qquad
-10 \overline{M}_{\text{pl}}^{2} m_{\phi}^{2} T\,
K_{1}^{2}\!\left(\frac{m_{\phi}}{T}\right)
\Biggr].
\end{align}
Here $R_{\text{portal}}(T)$ arises from the Higgs-portal interaction,
$R_{\text{grav}}(T)$ collects purely gravitational contributions
from the Higgs sector and other SM particles,
and $R_{\text{int}}(T)$ represents the interference between the Higgs-portal
and gravitational amplitudes.

Integrating Eq.~\eqref{DMeq2} over temperature from $T_{\mathrm{reh}}$ down to $T_{*}$,
with the hierarchy $T_{*} \ll m_\phi \ll T_{\mathrm{reh}}$,
we obtain a late-time DM yield that no longer depends on the choice of $T_{*}$.
The resulting asymptotic abundance can be written as
\begin{align}
Y(T_{*})
\simeq
Y(T_{\mathrm{reh}})
+
\lambda_{h\phi}^{2}\frac{9\sqrt{10}\,\overline{M}_{\text{pl}}}
{4096\,\pi^{4}\,g_{\text{reh}}^{1/2}\,m_\phi}
+
\frac{73\sqrt{10}\,T_{\text{reh}}^{3}}
{80\,\pi^{6}\,g_{\text{reh}}^{1/2}\,\overline{M}_{\text{pl}}^{3}}
+
\lambda_{h\phi}\frac{\sqrt{10}\,T_{\text{reh}}}
{16\,\pi^{6}\,g_{\text{reh}}^{1/2}\,\overline{M}_{\text{pl}}}.
\label{Y_T*}
\end{align}
In this expression, $g_{\mathrm{reh}}$ has been taken to be constant throughout the integration.
The value of the yield at the reheating temperature, $Y(T_{\mathrm{reh}})$,
is sensitive to the initial conditions and to the detailed dynamics during the reheating era and we put $Y(T_{\mathrm{reh}})=0$ in our analysis.

Using the asymptotic value of the yield in Eq.~\eqref{Y_T*},
the present-day DM relic abundance can be evaluated as
\begin{multline}
\Omega_\phi h^{2}
=
1.6\times 10^{8}
\left(
\frac{m_{\phi}}{1\,\mathrm{GeV}}
\right)
\left(
\frac{g_{0}}{g_{\mathrm{reh}}}
\right)
\Biggl[
\lambda_{h\phi}^{2}
\frac{9\sqrt{10}\,\overline{M}_{\text{pl}}}
{4096\,\pi^{4}\,g_{\text{reh}}^{1/2}\,m_{\phi}}
\\
+
\frac{73\sqrt{10}\,T_{\mathrm{reh}}^{3}}
{80\,\pi^{6}\,g_{\text{reh}}^{1/2}\,\overline{M}_{\text{pl}}^{3}}
+
\lambda_{h\phi}
\frac{\sqrt{10}\,T_{\mathrm{reh}}}
{16\,\pi^{6}\,g_{\text{reh}}^{1/2}\,\overline{M}_{\text{pl}}}
\Biggr].
\label{eq:Omegah^2}
\end{multline}
Here $g_{0}=3.91$ denotes the effective number of relativistic degrees of freedom at the present epoch.
The first term in Eq.~\eqref{eq:Omegah^2} corresponds to
IR-dominated freeze-in via the Higgs-portal interaction
and is controlled by temperatures of order $T \sim m_\phi$.
The second term represents UV-dominated freeze-in due to
graviton exchange and is governed by the highest temperature,
leading to a characteristic $T_{\mathrm{reh}}^{3}$ dependence.
The third term arises from the interference between the Higgs-portal
and gravitational amplitudes, yielding a linear dependence on
$T_{\mathrm{reh}}$.

\begin{figure}[t]
  \begin{center}
  \includegraphics[width=90mm]{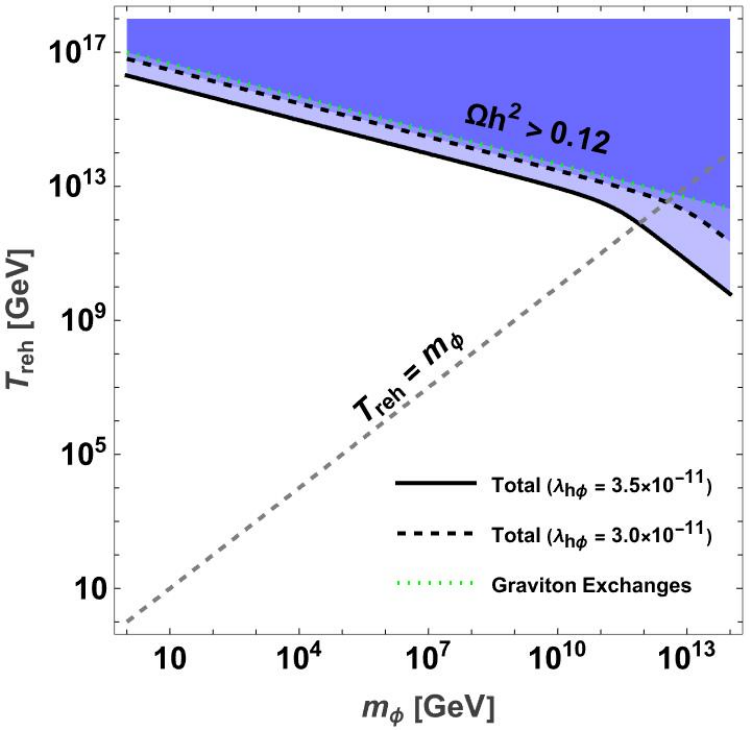}
  \end{center}
 \caption{Parameter space in $(m_\phi, T_{\mathrm{reh}})$ for the DM abundance.
  The blue shaded region indicates $\Omega h^2 > 0.12$.
  The region with $T_{\mathrm{reh}} > m_\phi$ in the figure
  (namely, the region above the gray dashed line)
  is valid under the current approximation~\eqref{Y_T*}.}
  \label{fig:RelicD}
\end{figure}
In Figure~\ref{fig:RelicD}, we show the parameter space in $(m_\phi, T_{\mathrm{reh}})$
under the condition that the observed DM abundance is saturated, $\Omega h^2 = 0.12$.
The green dotted line in Figure~\ref{fig:RelicD} represents the contribution
from gravitational interactions (corresponding to the second term in
Eq.~\eqref{eq:Omegah^2}),
whereas the black dashed and solid lines show the total DM production,
including the Higgs-portal interaction
(from the first and third terms in Eq.~\eqref{eq:Omegah^2}),
for $\lambda_{h\phi} = 3.0 \times 10^{-11}$ (black dashed)
and $\lambda_{h\phi} = 3.5 \times 10^{-11}$ (black solid), respectively.
The blue shaded region indicates DM overclosure, namely $\Omega h^2 > 0.12$.
We note that only the region with $T_{\mathrm{reh}} \gg m_\phi$ in the figure
(i.e., above the gray dashed line)
is consistent with the assumption used in Eq.~\eqref{Y_T*}.

Therefore, Higgs-portal scalar DM over a broad mass range
can account for the entire observed DM abundance
by varying the reheating temperature.
However, in order to avoid DM overproduction,
the Higgs-portal coupling must be tiny,
$\lambda_{h\phi} \lesssim 3.5 \times 10^{-11}$.
We also note that for $\lambda_{h\phi} = 3.6 \times 10^{-11}$,
the first term in Eq.~\eqref{eq:Omegah^2}
yields $\Omega h^2 \simeq 0.13$,
which exceeds the observed value.
When only the Higgs-portal interaction is taken into account,
the DM abundance is independent of both the DM mass
and the reheating temperature.

From Sec.~\ref{sec:posi_bound},
DM masses of order $10^{10}$--$10^{11}\,\mathrm{GeV}$ (or larger)
are required to realize a GUT-scale cutoff.
Consequently, reheating temperatures,
$T_{\mathrm{reh}} \lesssim 10^{14}\,\mathrm{GeV}$,
are required in Figure~\ref{fig:RelicD}.


\begin{figure}[t]
  \begin{center}
  \includegraphics[width=100mm]{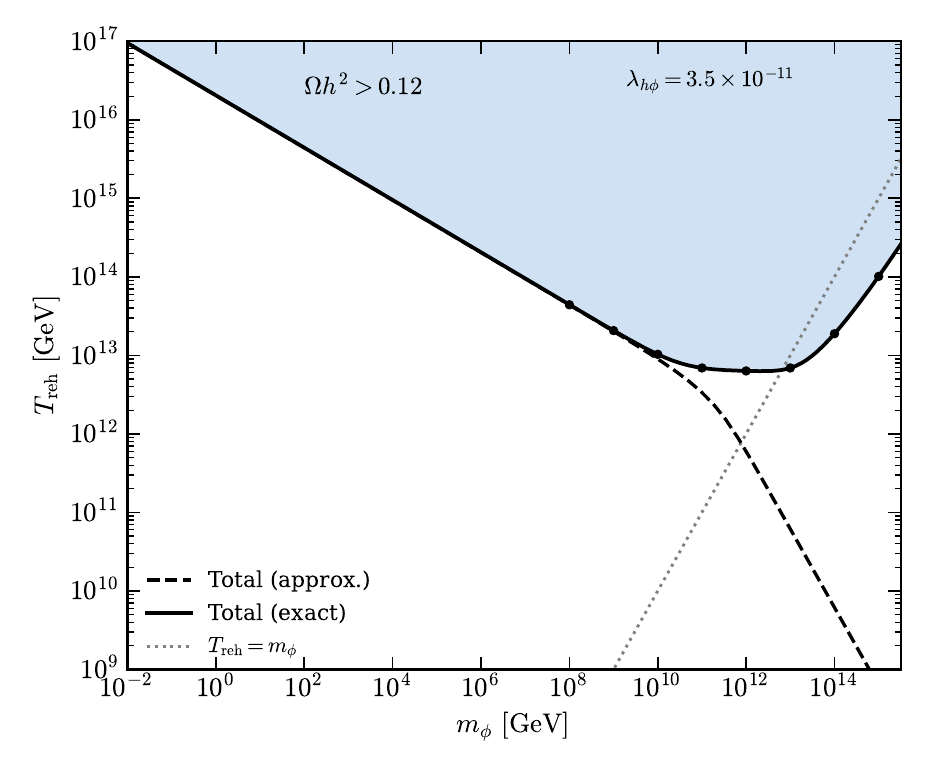}
  \end{center}
 \caption{Parameter space in $(m_\phi, T_{\rm reh})$ satisfying
the observed DM relic abundance, $\Omega_\phi h^2=0.12$, for
$\lambda_{h\phi}=3.5\times10^{-11}$. The dashed line shows the
approximate result of Eq.~\eqref{eq:Omegah^2}, valid for $T_{\rm reh}\gg m_\phi$.
The solid line shows the exact result of
Eqs.~\eqref{eq:Yexact} and \eqref{eq:Omegaexact} in Appendix~\ref{sec:appendix_C}, with circles marking
representative values of $m_\phi$ at which it was evaluated. The dotted
line indicates $T_{\rm reh}=m_\phi$. The shaded region indicates
$\Omega_\phi h^2>0.12$ (overproduction), evaluated using the exact
result.
}
  \label{fig:RelicD2}
\end{figure}
The bound $T_{\rm reh}\lesssim10^{14}\,{\rm GeV}$ derived above relies on
the approximate closed-form expression in Eq.~\eqref{Y_T*}, obtained under the
assumption $T_*\ll m_\phi\ll T_{\rm reh}$. To assess the validity of this
approximation, we have also evaluated the exact result,
Eqs.~\eqref{eq:Yexact}--\eqref{eq:Omegaexact} in Appendix~\ref{sec:appendix_C}, 
obtained by integrating
Eq.~\eqref{DMeq2} without assuming $T_{\rm reh}\gg m_\phi$. 
The resulting boundary of the $\Omega_\phi h^2=0.12$ contour is shown in Figure~\ref{fig:RelicD2}, together with the approximate result of Eq.~\eqref{eq:Omegah^2}, for $m_\phi$ up to
$10^{15}\,{\rm GeV}$, an order of magnitude below the GUT scale.
Importantly, the peak value
$T_{\rm reh}\lesssim10^{14}\,{\rm GeV}$, attained near
$m_\phi\sim10^{7\text{-}9}\,{\rm GeV}$ where the approximation is
excellent, remains an accurate upper bound across the full mass range
required by the gravitational positivity bounds discussed in
Sec.~\ref{sec:posi_bound}, up to $m_\phi\sim10^{15}\,{\rm GeV}$.

Note that this bound is remarkably insensitive to the precise value of
$\lambda_{h\phi}$. Near $m_\phi\sim10^{15}\,{\rm GeV}$, where the peak
value of $T_{\rm reh}$ in the exact treatment is attained, one has
$T_{\rm reh}\ll m_\phi$, so that every channel contributing to
the reaction rate -- Higgs-portal, gravitational, and their interference
-- is Boltzmann suppressed. However, the purely gravitational channel
(independent of $\lambda_{h\phi}$) carries a coefficient that grows with
a much higher power of $m_\phi$ than the $\lambda_{h\phi}$-dependent
channels, so that despite comparable suppression it numerically
dominates the total rate by several orders of magnitude near
$m_\phi\sim10^{15}\,{\rm GeV}$. As a result, lowering
$\lambda_{h\phi}$ from $3.5\times10^{-11}$ all the way to $0$ changes the
resulting bound at $m_\phi=10^{15}\,{\rm GeV}$ by only about one part in
$10^{5}$, from $T_{\rm reh}\simeq1.01269\times10^{14}\,{\rm GeV}$ to
$T_{\rm reh}\simeq1.01268\times10^{14}\,{\rm GeV}$; the slight decrease
reflects the negative sign of the interference term in this regime. We
therefore conclude that the bound $T_{\rm reh}\lesssim10^{14}\,{\rm GeV}$
is essentially independent of $\lambda_{h\phi}$, as long as it remains
small enough to avoid overproduction via the Higgs-portal channel alone.


In this analysis, we set the initial DM abundance
(the amount of DM produced at the end of the reheating epoch)
to zero, i.e., $Y(T_{\mathrm{reh}})=0$ in Eq.~\eqref{Y_T*}.
This initial condition depends on the inflaton dynamics during reheating.
However, since a nonzero initial abundance only increases
the final DM relic abundance,
the bound $T_{\mathrm{reh}} \lesssim 10^{14}\,\mathrm{GeV}$
remains valid, as a higher reheating temperature enhances
DM production from graviton-mediated interactions.

In the Higgs-$R^2$ inflation model studied in Ref.~\cite{Aoki:2022dzd}, the
reheating temperature can reach up to $T_{\rm reh}\sim10^{14}\,{\rm GeV}$,
dominated by scattering off the oscillating Higgs condensate, which is
consistent with the bound obtained here. In that model, the observed relic
abundance was found to be reproduced for DM masses up to
$m_\phi\sim10^{9}\,{\rm GeV}$, since a sizable fraction of the DM is
produced non-thermally during reheating, a channel neglected in the present
analysis (which sets $Y(T_{\rm reh})=0$). We note, however, that this
result of Ref.~\cite{Aoki:2022dzd} relies on the approximation
$T_{\rm reh}\gg m_\phi$ for the thermal production. Whether the Higgs-$R^2$
inflation model can accommodate the masses
$m_\phi\gtrsim10^{10\text{-}11}\,{\rm GeV}$ required by the gravitational
positivity bounds of Sec.~\ref{sec:posi_bound} therefore requires a
treatment beyond this approximation, which we leave for future work.
Similarly to Figure~\ref{fig:RelicD2}, larger DM masses in the
Higgs-$R^2$ inflation model may reproduce the observed relic abundance
with the reheating temperature of around $10^{14}\,{\rm GeV}$
predicted in the model.

\section{Summary and Discussion} \label{sec:summary}
We have considered a Higgs-portal real scalar DM model with a $Z_2$-odd parity and renormalizable interactions.  
In this framework, we applied gravitational positivity bounds to the forward DM scattering process 
$\phi \phi \to \phi \phi$.  

For light DM with $m_\phi < m_h$ and in the absence of a DM self-coupling, new physics beyond the Higgs-portal scenario is required to appear below an energy scale of $10^{10}$~GeV.
Extending the validity of the model to higher energies requires heavier DM, with masses around $10^{10}$~GeV allowing the cutoff scale to reach values comparable to the GUT scale or a typical string scale.
When a DM four-point self-coupling is included, achieving a cutoff scale at the GUT scale generally requires a heavier DM mass.
In particular, when the self-coupling dominates the positivity bound, the DM mass is required to be approximately one order of magnitude larger, around $10^{11}$~GeV.
We find that, in the presence of both interactions, achieving a cutoff scale at the GUT scale generally requires a DM mass of order $10^{10}$--$10^{11}\,\mathrm{GeV}$ (or above), with the larger values favored when the four-point self-coupling plays a significant role.

In this heavy-mass regime of Higgs-portal DM the observed relic abundance can be explained
by the freeze-in mechanism.
The Higgs-portal coupling must be tiny,
$\lambda_{h\phi} \lesssim 3.5 \times 10^{-11}$.
For DM masses of order
$10^{10}$--$10^{11}\,\mathrm{GeV}$ (or higher, up to $10^{15}\,\mathrm{GeV}$),
which are required to realize a GUT-scale cutoff,
the reheating temperature is constrained to be
$T_{\mathrm{reh}} \lesssim 10^{14}\,\mathrm{GeV}$.

We comment on the possibility of obtaining additional constraints from gravitational positivity.  
Processes suitable for positivity bounds must involve stable particles~\cite{Aoki:2022qbf,Aoki:2023tmu}.  
For the process $\phi e \to \phi e$, infrared (IR) divergences in $B_{\text{grav}}$ pose a significant challenge.  
While redefining asymptotic electron states using the Faddeev-Kulish formalism can resolve IR divergences in QED~\cite{Oller:2024qkw}, introducing a $t$-channel graviton exchange generates new contributions from diagrams with two photon propagators (and one electron propagator) in the triangle loop, leading to even more severe IR divergences from photons.

In the case of the process $\phi \gamma \to \phi \gamma$, the $B_{\text{non-grav}}$ contribution appears first at the two-loop level.  
In principle, this contribution can be computed via the optical theorem, providing a potential pathway for further refinement of the bounds.

\paragraph*{Acknowledgements}
%
We have used the package TikZ-Feynman~\cite{Ellis:2016jkw} to draw the Feynman diagrams.
We would like to express our sincere gratitude to Katsuki Aoki for useful discussions and comments, Kaoru Hagiwara for his encouragement, Akio Sugamoto for his valuable input on infrared divergences, Tatsu Takeuchi for carefully reading the manuscript and providing valuable feedback, and Kazuya Yonekura for his helpful comments on string theory.
We also thank the Astrophysics \& Cosmology Group at the Yukawa Institute for Theoretical Physics, Kyoto University, for their warm hospitality, where a part of this work was carried out. 
This work is supported by JSPS KAKENHI Grant Number JP24K17040.
%

\appendix

\section{Derivation from Eq.~\eqref{eq:amp1_s} to Eq.~\eqref{eq:amp2_s}} \label{sec:appendix_posi}
%
%
\begin{figure}[t]
\begin{center}
\includegraphics[width=0.7\textwidth,clip]{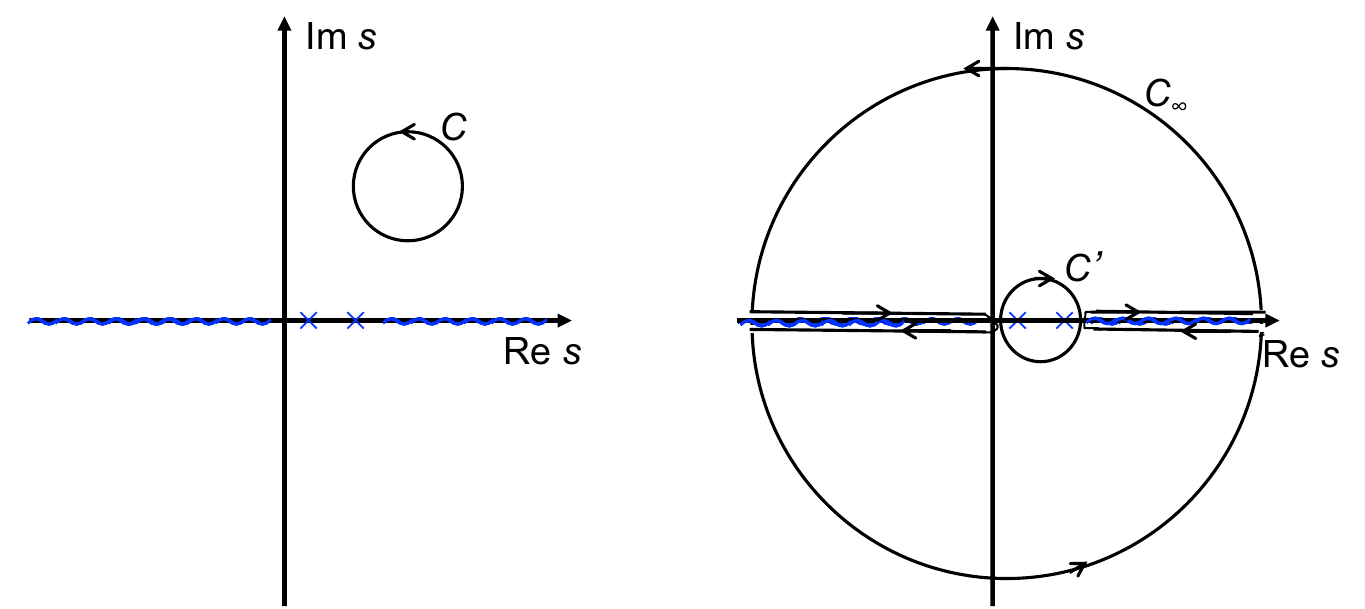}
\end{center}
\hspace{26mm}\includegraphics[width=0.4\textwidth,clip]{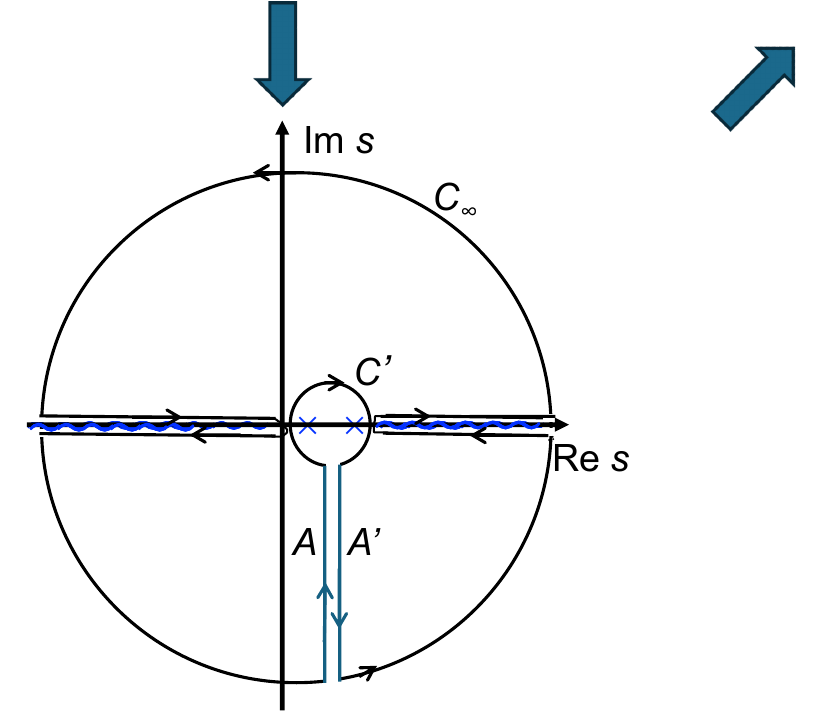}
\caption{Scattering amplitude $\mathcal{M}(s,t)$ in the complex $s$-plane with poles and branch cuts.
The integration contour $\mathcal{C}$ in the upper-left panel can be deformed into the contours $\mathcal{C}'$ and $\mathcal{C}_\infty$ in the upper-right panel, following the analytic structure of the amplitude aside from poles and branch cuts.
The bottom-left panel shows how this deformation is achieved by introducing two paths, $A$ and $A'$, which can be brought arbitrarily close to each other and ultimately removed.}
\label{fig:s_plane}
\end{figure}

From assumption (ii), valid above the energy scale $E>\Lambda$, the scattering amplitude obeys Lorentz invariance, analyticity (except for poles and branch cuts), crossing symmetry (from the LSZ reduction formula), and unitarity.
Using this, we employ a twice-subtracted dispersion relation, Eq.~\eqref{eq:amp2_s}, proportional to $\mathrm{Im}\,\mathcal{M}(s,t)/s^3$, to relate the real part of the amplitude, Eq.~\eqref{eq:amp3}.
This is justified by the high-energy behavior in Eq.~\eqref{eq:amp_s_limit} (assumption (i)) and is used to extract the leading contribution to the amplitude relevant for the gravitational positivity bounds.

Let us consider the (nearly) forward elastic amplitude $\mathcal{M}(s,t)$ in the complex $s$-plane, as shown in Figure~\ref{fig:s_plane}.
The contour $\mathcal{C}$ in Eq.~\eqref{eq:amp1_s} can be deformed into $\mathcal{C}'$ and $\mathcal{C}_\infty$.

With the contour $\mathcal{C}$ shown in the upper-left panel of Figure~\ref{fig:s_plane}, the amplitude can be written as
\begin{align}
\mathcal{M}(s,t) =
\left(s-2m_\phi^2+\frac{t}{2}\right)^2
\oint_{\mathcal{C}} \frac{ds'}{2\pi i}
\frac{\mathcal{M}(s',t)}{(s'-s)\left(s'-2m_\phi^2+t/2\right)^2},
\tag{\ref{eq:amp1_s}}
\end{align}
where the factor $\left(s-2m_\phi^2+t/2\right)^2$~\cite{Aoki:2023khq} ensures that the $u$-channel amplitude, arising from crossing symmetry, takes the same functional form.

Deforming $\mathcal{C}$ into $\mathcal{C}'$ and $\mathcal{C}_\infty$ as in Figure~\ref{fig:s_plane} leads to the twice-subtracted dispersion relation,
\begin{align}
\mathcal{M}(s,t)
&= \left[ \frac{a_{-1}(t)}{s-m_\phi^2} + (s\leftrightarrow u(s,t)) \right]
  + a_0 + a_1(t)s \nonumber \\
&\quad + 2 \frac{\bigl(\bar{s}+\bar{t}/2\bigr)^2}{\pi}
\int_{4m_\phi^2}^{\infty} d\mu\,
\frac{\mathrm{Im}\,\mathcal{M}(\mu + i\epsilon, t)}
     { \bigl(\bar{\mu} + \bar{t}/2\bigr)
       \bigl[ (\bar{\mu} + \bar{t}/2)^2 - (\bar{s} + \bar{t}/2)^2 \bigr]},
\tag{\ref{eq:amp2_s}}
\end{align}
where barred variables are defined by $\bar{z}=z-4m_\phi^2/3$.

The explicit forms of $a_{-1}(t)$, $a_0$, and $a_1(t)$ are not required for our analysis.
The bottom-left panel of Figure~\ref{fig:s_plane} illustrates how the upper-right panel is obtained by introducing two nearly coincident paths, $A$ and $A'$, which can eventually be shrunk to zero.
The contribution from $\mathcal{C}_\infty$ vanishes because of assumption (i) in the limit $|s|\to\infty$, as shown later in this appendix.

To outline the derivation of Eq.~\eqref{eq:amp2_s} from Eq.~\eqref{eq:amp1_s}, let us first consider a part of the contour 
$\mathcal{C}'$ contribution.
The $s-u$ crossing symmetry for the scalar field is used as
\begin{align}
\mathcal{M}(s, t) = \mathcal{M}(u, t).
\end{align}
With the on-shell condition $u = 4m^2_{\phi} - s - t$,
\begin{align}
\mathcal{M}(s, t) = \mathcal{M}(4m^2_{\phi} - s - t, t).
\end{align}
This implies that for $(0 <) -t \ll 4m^2_{\phi}$, when the amplitude has a pole at $s = m^2_{\phi}$,
it also has a pole at $s \sim 3 m^2_{\phi}$. 
Both originate from the tree-level forward elastic scattering
$\phi\phi \to \phi\phi$ mediated by $\phi$ in the $s$- and $u$-channels.
By the residue theorem,
\begin{align}
\oint_{\mathcal{C}'} \frac{ds'}{2\pi i} \frac{\mathcal{M}(s', t)}{(s'-s)\left(s'-2m^2_\phi+t/2\right)^2} &= 
- \frac{\mathrm{Res}_{s'=m^2_\phi}\mathcal{M}(s', t)}{(m^2_\phi - s)\left(-m^2_\phi+t/2\right)^2} + (\mathrm{crossing}).\label{eq:pole}
\end{align}
To obtain the relevant part of Eq.~\eqref{eq:amp1_s}, we multiply $(s-2m^2_\phi+t/2)^2$ to the above Eq.~\eqref{eq:pole}.
Here 
\begin{align}
(s-2m^2_\phi+t/2)^2 = (-m^2_\phi+t/2 - m^2_\phi + s)^2 = (m^2_\phi-t/2)^2(1-(s-m^2_\phi)/(m^2_\phi-t/2))^2.
\end{align}
Thus, multiplying Eq.~\eqref{eq:pole} by the factor $(s-2m^2_\phi+t/2)^2$ gives
%
\begin{align}
\left(s-2m^2_\phi+\frac{t}{2}\right)^2\oint_{\mathcal{C}'} \frac{ds'}{2\pi i} \frac{\mathcal{M}(s', t)}{(s'-s)\left(s'-2m^2_\phi+t/2\right)^2}
&=  \left[\frac{a_{-1}(t)}{s-m^2_\phi} + (s\leftrightarrow u(s,t))\right] \nonumber\\
&\quad+ a_0 + a_1(t)s, \label{eq:pole2}
\end{align}
with functions $a_{-1}(t)$, $a_0$, and  $a_1(t)$. 
Eq.~\eqref{eq:pole2} reproduces the first four terms in Eq.~\eqref{eq:amp2_s}, corresponding to pole contributions.
As noted in Sec.~\ref{sec:posi_review}, this pole contribution is generic to a scalar
field with a cubic self-interaction and is included here only as part
of the general pedagogical derivation of the dispersion relation; it is
forbidden by the $Z_2$ symmetry of the model in Eq.~\eqref{eq:lag2} and does not
appear in the explicit one-loop calculation of Sec.~\ref{sec:posi_bound}.

Next, we see the contribution from $\mathcal{C}_{\infty}$ vanishes.
\begin{align}
\left|\oint_{\mathcal{C}_{\infty}} ds'\frac{\mathcal{M}(s', t)}{(s'-s)(s'-2m^2_\phi+t/2)^2}\right|
&= \left|\oint_{\mathcal{C}_{\infty}} ds'\frac{\mathcal{M}(s', t)}{s'^3}\right| \nonumber \\
&\sim \lim_{|s'| \to \infty} \left|\frac{\mathcal{M}(s', t)}{s'^2}\right| = 0, \label{eq: inf_contour}
\end{align}
where the last step uses Eq.~\eqref{eq:amp_s_limit} in (i) with $t<0$.
The reason for dividing $\mathcal{M}(s, t)$ by $s^3$
is precisely to suppress this contribution and make the dispersion relation well-defined.
By doing so, we can ignore the $\mathcal{C}_{\infty}$ contribution.

The remaining contribution arises from the branch cuts in the complex s-plane. These appear when s reaches the on-shell energy of two $\phi$ particles, i.e. from one-loop intermediate states that generate an imaginary part of the amplitude. Explicitly, the relevant pieces are
\begin{align}
\left(\int_{4m^2_\phi}^{\infty} + \int_{-\infty}^0 \right) 
d\mu\frac{\mathcal{M}(\mu + i\epsilon, t)}{(\mu-s)(\mu - 2m^2_\phi + t/2)^2}\nonumber \\
+ \left(\int_{\infty}^{4m^2_\phi} + \int_0^{-\infty}\right)
d\mu\frac{\mathcal{M}(\mu - i\epsilon, t)}{(\mu-s)(\mu - 2m^2_\phi + t/2)^2}. \label{eq:branch}
\end{align}
Because of the on-shell condition $s + t + u = 4m^2_\phi$ with $-t \ll 4m^2_\phi$,
the branch cut for $s \geq 4m^2_\phi$ corresponds to the cut for $u \lesssim 0$.
By crossing symmetry, $\mathcal{M}(s, t) = \mathcal{M}(u, t)$,
another branch cut starts from $s=0$ to the negative $s$.
We introduce the shifted variables $\bar{z} = z - (4m^2_\phi/3)$, so that
$\bar{s} + \bar{t} + \bar{u} = 0$ from the on-shell condition.

The positive real $s$ parts of Eq.~\eqref{eq:branch} become
\begin{align}
&\int_{4m^2_\phi}^{\infty}d\mu\frac{\mathcal{M}(\mu + i\epsilon, t)}{(\mu-s)(\mu - 2m^2_\phi + t/2)^2}
+\int_{\infty}^{4m^2_\phi} d\mu\frac{\mathcal{M}(\mu - i\epsilon, t)}{(\mu-s)(\mu - 2m^2_\phi + t/2)^2}\nonumber \\
&\, \, \, \, = \int_{4m^2_\phi}^{\infty}d\mu\frac{\mathcal{M}(\mu + i\epsilon, t)}{(\mu-s)(\mu - 2m^2_\phi + t/2)^2}
- \int_{4m^2_\phi}^{\infty} d\mu\frac{\mathcal{M}(\mu - i\epsilon, t)}{(\mu-s)(\mu - 2m^2_\phi + t/2)^2}\nonumber \\
&\, \, \, \, = \int_{4m^2_\phi}^{\infty}d\mu\frac{\mathcal{M}(\mu + i\epsilon, t)}{(\mu-s)(\mu - 2m^2_\phi + t/2)^2}
- \int_{4m^2_\phi}^{\infty} d\mu\frac{\mathcal{M}^*(\mu + i\epsilon, t)}{(\mu-s)(\mu - 2m^2_\phi + t/2)^2}\nonumber \\
&\, \, \, \, = 2i \int_{4m^2_\phi}^{\infty}d\mu\frac{\mathrm{Im}\mathcal{M}(\mu + i\epsilon, t)}{(\mu-s)(\mu - 2m^2_\phi + t/2)^2}.\label{eq:posi}
\end{align}
We use the Schwarz reflection principle $\mathcal{M}(s^*,t) = \mathcal{M}^*(s,t)$ from the Analyticity of the amplitude 
from the second to the third lines.
Similarly, the negative-real-$s$ part in Eq.~\eqref{eq:branch} is
\begin{align}
& \int_{-\infty}^0 d\mu\frac{\mathcal{M}(\mu + i\epsilon, t)}{(\mu-s)(\mu - 2m^2_\phi + t/2)^2}
+ \int_0^{-\infty}d\mu\frac{\mathcal{M}(\mu - i\epsilon, t)}{(\mu-s)(\mu - 2m^2_\phi + t/2)^2}\nonumber \\
&\, \, \, \, = - \int_0^{-\infty} d\mu\frac{\mathcal{M}(\mu + i\epsilon, t)}{(\mu-s)(\mu - 2m^2_\phi + t/2)^2}
+ \int_0^{\infty}d\mu\frac{\mathcal{M}(-\mu - i\epsilon, t)}{(\mu+s)(\mu + 2m^2_\phi - t/2)^2}\nonumber \\
&\, \, \, \, = -\int_0^{\infty} d\mu\frac{\mathcal{M}(-\mu + i\epsilon, t)}{(\mu+s)(\mu + 2m^2_\phi - t/2)^2}
+ \int_0^{\infty}d\mu\frac{\mathcal{M}(-\mu - i\epsilon, t)}{(\mu+s)(\mu + 2m^2_\phi - t/2)^2}\nonumber \\
&\, \, \, \, = -\int_0^{\infty} d\mu\frac{\mathcal{M}(4m^2_\phi + \mu - i\epsilon-t, t)}{(\mu+s)(\mu + 2m^2_\phi - t/2)^2}
+ \int_0^{\infty}d\mu\frac{\mathcal{M}(4m^2_\phi + \mu + i\epsilon-t, t)}{(\mu+s)(\mu + 2m^2_\phi - t/2)^2}\nonumber \\
&\, \, \, \, = \int_{4m^2_\phi-t}^{\infty}d\mu\frac{\mathcal{M}(\mu + i\epsilon, t)}{(\mu+t - 4m^2_\phi+s)(\mu - 2m^2_\phi + t/2)^2}\nonumber \\
&\, \, \, \, \, \, \, \, - \int_{4m^2_\phi-t}^{\infty} d\mu\frac{\mathcal{M}(\mu - i\epsilon, t)}{(\mu+t - 4m^2_\phi+s)(\mu - 2m^2_\phi + t/2)^2}\nonumber \\
&\, \, \, \, = 2i\int_{4m^2_\phi}^{\infty}d\mu\frac{\mathrm{Im}\mathcal{M}(\mu + i\epsilon, t)}{(\mu+t - 4m^2_\phi+s)(\mu - 2m^2_\phi + t/2)^2}.\label{eq:neg}
\end{align}
Here we employed crossing symmetry and the on-shell condition,
e.g.,
$\mathcal{M}(-s + i\epsilon) = \mathcal{M}(4m^2_\phi -(-s+ i\epsilon) -t) =  \mathcal{M}(4m^2_\phi + s - i\epsilon -t)$
and 
$\mathcal{M}(-s - i\epsilon) = \mathcal{M}(4m^2_\phi -(-s- i\epsilon) -t) =  \mathcal{M}(4m^2_\phi + s + i\epsilon -t)$.
Additionally, we changed the integration variable from $\mu$ to $\mu - 4m^2_\phi + t$,
used the condition of $-t \ll 4m^2_\phi$, and applied the Schwarz reflection principle.

Adding Eqs.~\eqref{eq:posi} and \eqref{eq:neg} gives
\begin{align}
&2i\int_{4m^2_\phi}^{\infty}d\mu\frac{\mathrm{Im}\mathcal{M}(\mu + i\epsilon, t)}{(\mu - 2m^2_\phi + t/2)^2}
\left(\frac{1}{\mu-s}+\frac{1}{(\mu+t - 4m^2_\phi+s)}\right)\nonumber\\
&\, \, \, \, = 4i\int_{4m^2_\phi}^{\infty}d\mu\frac{\mathrm{Im}\mathcal{M}(\mu + i\epsilon, t)}{\left(\bar{\mu} + \bar{t}/2 \right)\left[\left(\bar{\mu} + \bar{t}/2 \right)^2 - \left(\bar{s} + \bar{t}/2\right)^2\right]}.\label{eq:fin}
\end{align}
By multiplying Eq.~\eqref{eq:fin} by $(s - 2m_\phi^{2} + t/2)^2 = (\bar{s} + \bar{t}/2)^2$ and dividing it by $2\pi i$, we obtain
\begin{align}
2\frac{\left(\bar{s}
+\bar{t}/2\right)^2}{\pi} \int_{4m^2_\phi}^{\infty}d\mu\frac{\mathrm{Im}\mathcal{M}(\mu + i\epsilon, t)}{\left(\bar{\mu} + \bar{t}/2 \right)\left[\left(\bar{\mu} + \bar{t}/2 \right)^2 - \left(\bar{s} + \bar{t}/2\right)^2\right]},
\end{align}
which corresponds to the integrated part in Eq.~\eqref{eq:amp2_s}.
%
%

\section{Each contribution to $B^{\phi \phi \to \phi \phi}_{\text{grav}}$} \label{sec:appendix_B}

We show explicit forms of each contribution to $B^{\phi \phi \to \phi \phi}_{\text{grav}}$ in Eq.~\eqref{eq:grav_B}, 
i.e., $B^{\phi \phi \to \phi \phi}_{\text{grav(a/b/c/d)}}$, corresponding to (a)--(d) in Figure~\ref{fig:diagram_grav}, respectively, in this Appendix.
\begin{align}
B^{\phi \phi \to \phi \phi}_{\text{grav(a)}} &= -\frac{\lambda^2_{h\phi} v^2}{96 \pi^2\overline{M}^2_{\text{pl}} m^4_h} f_{(a)}(m_{\phi}/m_h), \\
f_{(a)}(x) &= \frac{1}{x^8(1 - 4x^2)^2}\biggl\{120 x^8-154 x^6+55 x^4-6 x^2-(1-4 x^2)^2(3 x^4-8 x^2+3) \ln(x^2)\nonumber\\
&\, \, \, \, \, \, - 2\sqrt{1-4 x^2}(4x^8-54 x^6+69 x^4-26 x^2+3)\ln\left[\frac{1}{2x}\left(\sqrt{1-4x^2}+1\right)\right]\biggr\},
\end{align}
\begin{align}
B^{\phi \phi \to \phi \phi}_{\text{grav(b)}} &= -\frac{\lambda^2_{h\phi} v^2}{96 \pi^2\overline{M}^2_{\text{pl}} m^4_h} f_{(b)}(m_{\phi}/m_h), \\
f_{(b)}(x) &= \frac{1}{x^8(1 - 4x^2)^2}\biggl\{-64 x^8+124 x^6-51x^4+6x^2+(1-4 x^2)^2(x^4-6 x^2+3) \ln(x^2)\nonumber\\
&\, \, \, \, \, \, + 2\sqrt{1-4 x^2}(-30 x^6+55 x^4-24 x^2+3)\ln\left[\frac{1}{2x}\left(\sqrt{1-4x^2}+1\right)\right]\biggr\},
\end{align}
\begin{align}
B^{\phi \phi \to \phi \phi}_{\text{grav(c)}} &= B^{\phi \phi \to \phi \phi}_{\text{grav(a)}},\, \, \, \, B^{\phi \phi \to \phi \phi}_{\text{grav(d)}} = B^{\phi \phi \to \phi \phi}_{\text{grav(b)}}.
\end{align}

\section{DM Relic Abundance Beyond the Approximation $m_\phi \ll T_{\rm reh}$}\label{sec:appendix_C}
Let $x \equiv m_\phi/T$, and define the Meijer-$G$ functions
\begin{align}
G_a(x) &\equiv G^{3,0}_{1,3}\!\left(\genfrac{}{}{0pt}{}{1}{-\frac32,\ \frac32,\ \frac52}\,\bigg|\,x^2\right), \\[4pt]
G_b(x) &\equiv G^{3,1}_{2,4}\!\left(\genfrac{}{}{0pt}{}{1,\,1}{-\frac12,\ \frac32,\ \frac52,\ 0}\,\bigg|\,x^2\right), \\[4pt]
G_c(x) &\equiv G^{3,1}_{2,4}\!\left(\genfrac{}{}{0pt}{}{1,\,2}{-\frac32,\ \frac32,\ \frac52,\ 0}\,\bigg|\,x^2\right), \\[4pt]
G_d(x) &\equiv G^{3,1}_{2,4}\!\left(\genfrac{}{}{0pt}{}{1,\,2}{\ \ \frac12,\ \frac32,\ \frac52,\ 0}\,\bigg|\,x^2\right),
\end{align}
whose limits for $x\to\infty$ (i.e.\ $T_*\to0$) are $G_a(\infty)=0$ and
\begin{equation}
G_b(\infty) = -\frac34\pi^{3/2}, \qquad
G_c(\infty) = \frac12\pi^{3/2}, \qquad
G_d(\infty) = \frac38\pi^{3/2}.
\end{equation}
Integrating the Boltzmann equation~\eqref{DMeq2} over temperature from $T_{\rm reh}$ down to $T_*$, without assuming
$T_{\rm reh}\gg m_\phi$, and taking the $T_*\to0$ limit (which we find
converges already for $T_*/m_\phi\lesssim1/20$, i.e.\ well before the
reaction rate becomes numerically negligible), the DM yield at
late times is given in closed form by
the following expression, where $G_i(x)$ is evaluated at $x=m_\phi/T_{\rm reh}$:
\begin{align}
Y(T_{\rm reh}) = \frac{1}{512\sqrt{10}\,\pi^{11/2}\,\overline M_{\rm pl}^3\,m_\phi\,\sqrt{g_{\rm reh}}}
\Big\{ \, & 16\, m_\phi^4\, G_a(x) \notag\\
+\, & 8\, m_\phi^2\big(m_\phi^2-5\overline M_{\rm pl}^2\lambda_{h\phi}\big)\big[G_b(\infty)-G_b(x)\big] \notag\\
+\, & 1035\, m_\phi^4\big[G_c(\infty)-G_c(x)\big] \notag \\
+\, & 2\big(3m_\phi^4-10\overline M_{\rm pl}^2 m_\phi^2\lambda_{h\phi}+15\overline M_{\rm pl}^4\lambda_{h\phi}^2\big) \notag\\
&\quad\times\big[G_d(\infty)-G_d(x)\big] \Big\},
\label{eq:Yexact}
\end{align}
and the DM relic abundance can be evaluated as
\begin{equation}
\Omega_\phi h^2 = 1.6\times10^{8}\left(\frac{m_\phi}{1\,{\rm GeV}}\right)\!\left(\frac{g_0}{g_{\rm reh}}\right) Y(T_{\rm reh}).
\label{eq:Omegaexact}
\end{equation}
Equations~\eqref{eq:Yexact}--\eqref{eq:Omegaexact} are exact: unlike
Eq.~\eqref{Y_T*}, they involve no assumption relating $T_{\rm reh}$ and $m_\phi$,
and reduce to Eqs.~\eqref{Y_T*} and \eqref{eq:Omegah^2}, respectively, in the limit $x=m_\phi/T_{\rm reh}\to0$.



\end{document}